\documentclass[aps,floats]{revtex4}
\usepackage{tabularx,graphicx}
\usepackage{epsfig}
\usepackage{amsmath}
\usepackage{bbm}
\usepackage{graphicx}

\begin{document}
%\draft
%\flushbottom
%\twocolumn[
%\hsize\textwidth\columnwidth\hsize\csname @twocolumnfalse\endcsname

\title{Magnetic and Kohn-Luttinger instabilities near a Van Hove 
singularity: monolayer versus twisted bilayer graphene.}
\author{J. Gonz\'alez  \\}
\address{Instituto de Estructura de la Materia,
        Consejo Superior de Investigaciones Cient\'{\i}ficas, Serrano 123,
        28006 Madrid, Spain}

\date{\today}

\begin{abstract}
%\widetext

We investigate the many-body instabilities of electrons 
interacting near Van Hove singularities arising in monolayer and twisted 
bilayer graphene.
We show that a pairing instability must be dominant over the tendency 
to magnetic order as the Fermi level is tuned to the Van Hove singularity in 
the conduction band of graphene. As a result of the extended character of 
the saddle points in the dispersion, we find that the pairing of the 
electrons takes place preferentially in a channel of $f$-wave symmetry, 
with an order parameter vanishing at the position of the saddle points 
along the Fermi line. In the case of the twisted bilayers, the dispersion 
has instead its symmetry reduced down to the $C_{3v}$ group and, most 
importantly, it leads to susceptibilities that diverge at the saddle points 
but are integrable along the Fermi line. This implies that a ferromagnetic 
instability becomes dominant in the twisted graphene bilayers near the Van
Hove singularity, with a strength which is amplified as the lowest subband 
of the electron system becomes flatter for decreasing twist angle.

\end{abstract}
%\pacs{71.10.Pm,73.22.-f}

%]
\maketitle

\section{Introduction}

In recent years, there has been great interest in the investigation of the 
correlations that may arise from $e$-$e$ interactions in 
monolayer\cite{khves,her,jur,drut1,hands,gama,fer,ggg,prb} 
as well as in bilayer graphene\cite{vaf,zhang,nan,lemo}. 
In these systems, the Coulomb repulsion between 
electrons constitutes the dominant interaction, placing the carbon material in 
a strong coupling regime as $e^2$ turns out to be nominally larger than the 
Fermi velocity $v_F$ of the electrons\cite{kot}. Yet the effects of electron 
correlations have been quite elusive, apart from the observation 
of the fractional quantum Hall effect in monolayer graphene\cite{and,bol}
and several signatures of exotic phases in bilayer graphene\cite{may,vel}.

The observation of superconductivity in a graphene system is a widely shared 
aspiration, and some proposals have been already put forward to induce a pairing 
instability in the carbon layer\cite{gon,uch,black,hon,prbkl,paco,barlas}. 
Several experimental studies have been carried out showing the feasibility of 
superconducting correlations and even supercurrents in graphene contacted with 
superconducting electrodes\cite{heer,shai}. However, it is still an intriguing 
question whether graphene may support superconducting correlations on its own 
under appropriate experimental conditions.

In that respect, a suitable way of amplifying the electronic correlations may
consist in tuning the Fermi level at the points with divergent density of 
states (so-called Van Hove singularities) that are present in the spectrum of
both monolayer graphene and twisted graphene bilayers. Experimental measures of 
the electronic dispersion in graphene at large doping levels have shown indeed 
that the saddle points in the conduction band develop an extended shape that 
may significantly reinforce the modulation of the screened Coulomb 
interaction\cite{prl}. On the other hand, a Van Hove singularity has been 
already observed experimentally in the lowest-energy subband of the twisted 
bilayers\cite{li}, which makes them ideal systems to address the effects of the 
strong correlation with a minimum of electron doping.

In narrow-band electron systems, magnetic and superconducting instabilities 
are in general likely to appear as a consequence of the enhanced density of states 
near the Fermi level. The route towards superconductivity can be elaborated  
starting from ideas proposed long ago by W. Kohn and J. M. Luttinger, trying
to understand if a pairing instability could arise out of a purely repulsive 
interaction\cite{kl}. It happens that, in electron systems with anisotropic
dispersion, the Coulomb interaction can be screened with different 
intensity along the Fermi surface, giving rise in some cases to a significant 
modulation of the effective interaction. Then it becomes possible that, after 
making the decomposition into the different modes according to the symmetry 
group of the Fermi surface, the couplings in some of the channels may turn out 
to be negative\cite{bar,prl2}. This sign of attractive interaction is enough to 
trigger a superconducting instability, though the magnitude of the negative 
couplings may be in general so small that the critical scale for 
superconductivity becomes many orders of magnitude smaller than the Fermi 
energy.

However, in cases where the Fermi surface is close to saddle points in the 
electronic dispersion, the modulation displayed by the screened 
$e$-$e$ interaction can be quite strong, as a result of the divergent 
density of states provided by the saddle points. In models with such a Van Hove 
singularity in the spectrum, the critical scale for the pairing instability may 
be a small fraction of the typical energy scale of the band structure, 
leading nevertheless to much higher transition temperatures than those obtained 
in conventional models with electron-phonon interactions. In the context of the 
high-$T_c$ cuprate superconductors, the proximity of the Fermi level to a 
saddle point in the electronic dispersion\cite{gof,shen} has been invoked to 
account for several unconventional properties of the cuprates\cite{newns,mark}, 
including the $d$-wave order parameter of the superconducting 
condensate\cite{epl,zan}. Many of these features enter also in correspondence 
with the properties expected from microscopic theories of the cuprates, based 
mainly on the Hubbard model.

In the context of the honeycomb carbon lattice, an investigation of the 
role of the Van Hove singularities to induce a superconducting instability has 
been carried out in Ref. \cite{prbkl}. The analysis was made there under 
the assumption of a relatively small next-to-nearest neighbor hopping, still
preserving an approximate nesting of the Fermi line passing by the 
saddle points. Under these conditions, the electron scattering is more intense 
at the momentum connecting each two inequivalent saddle points, and the 
superconducting instability turns out to appear in the channel with $d$-wave 
symmetry. A similar conclusion was reached in the analysis of Ref. \cite{nl}. 
Anyhow, that geometry of the Fermi line does not seem to be applicable to the 
Van Hove singularity in the conduction band of graphene, where the extended 
character of the saddle points favors the scattering of electrons with 
vanishing momentum transfer. Then, the development of the superconducting 
instability takes place in general in the $f$-wave channel\cite{prl}. The same 
symmetry of the order parameter has been also obtained in numerical studies of 
the Van Hove singularity under the assumption of short-range Coulomb 
interaction\cite{thom}. In a different unrelated situation, $f$-wave symmetry 
has been also predicted for a superconducting instability arising in graphene 
at low doping levels about the charge neutrality point\cite{hon}.

In the present paper, we will apply the Kohn-Luttinger mechanism of 
superconductivity to the case where there is an extended Van Hove singularity 
in the electronic spectrum, as it happens in the conduction band of graphene. 
We will provide a very general argument to show that the superconducting 
instability must have then $f$-wave symmetry. As the electron scattering is 
greatly enhanced by the extended character of the saddle points in the 
dispersion, the uniform magnetic susceptibility can also grow large in that 
situation. We will see however that the couplings measuring the effective 
attraction in the pairing channels are also amplified, formally 
diverging when the critical point for a ferromagnetic instability is 
approached. This explains that the pairing instability turns out to prevail 
over the tendency towards magnetic order in the presence of the extended Van 
Hove singularity.

In the case of the twisted graphene bilayers,  we will see instead that they
are more prone to develop a magnetic instability as they approach the regime
close to the formation of flat zero-energy subbands for decreasing twist 
angle\cite{port,bist,prl3}. We will show that the low-energy Van Hove 
singularity arising from the hybridization of twisted Dirac cones favors 
then a ferromagnetic instability in the system. This is reminiscent of the 
evidence of ferromagnetism found in Monte Carlo simulations of the Hubbard 
model in the square lattice with next-to-nearest neighbor hopping, in the 
limit where the Van Hove singularity undergoes a similar collapse at the 
bottom of the band\cite{sor}. As shown below, the many-body approaches used 
here to describe these instabilities can be also put in correspondence with 
renormalization group analyses of the interacting electron system, which have 
also shown that ferromagnetism is a likely instability when the Fermi level 
is close to a Van Hove singularity, in the absence of significant nesting 
of the Fermi surface\cite{ferro}.

\section{Magnetic and pairing instabilities}

\begin{figure}
\begin{center}
\mbox{\epsfxsize 7cm \epsfbox{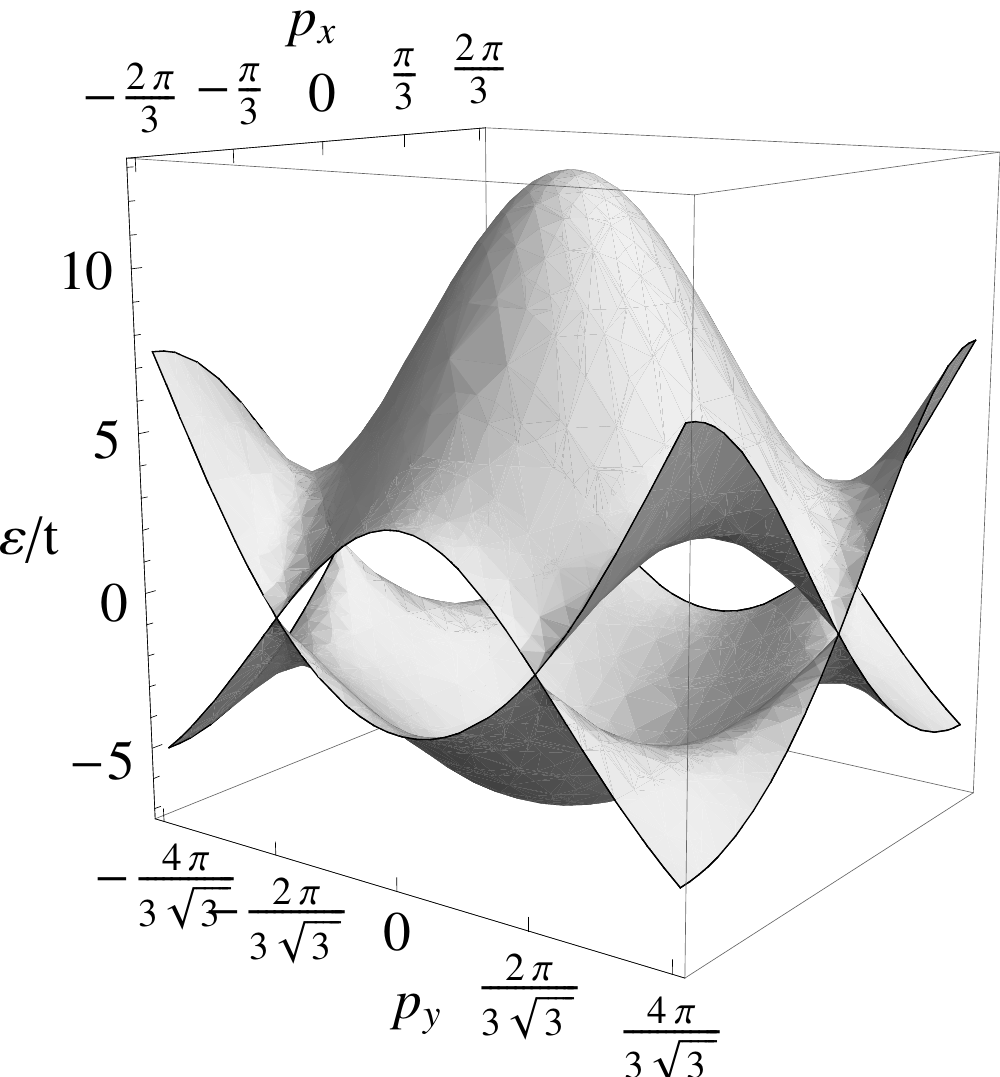}}
\end{center}
\caption{Plot of the dispersion of the conduction and valence bands from a 
tight-binding model of graphene (energy is measured in units of the 
nearest-neighbor hopping $t$ and momentum in units of the inverse of the 
C-C distance).}
\label{disp}
\end{figure}

In a two-dimensional system, the Van Hove singularities in the density of 
states arise from the presence of saddle points in the electronic dispersion, 
as those shown in a typical plot for the graphene lattice in 
Fig. \ref{disp}. For small deviation ${\bf k}$ of the momentum with
respect to the center of each saddle point, the dispersion can be approximated
by 
\begin{equation}
\varepsilon ({\bf k}) \approx 
                        \alpha \:  k_x^2 - \beta \:  k_y^2 
\label{sp}
\end{equation}

As a consequence of the flatness of the band around the saddle points, a number
of susceptibilities diverge at the Van Hove singularity, with relative 
strengths that depend on the particular values of the $\alpha $ and $\beta $
parameters. The particle-hole susceptibility $\chi_{\rm ph} ({\bf q}, \omega )$ 
at vanishing momentum transfer is given for instance by
\begin{equation}
\chi_{\rm ph} ({\bf 0}, \omega ) \approx \frac{1}{4 \pi^2} 
     \frac{1}{\sqrt{\alpha \beta }}  
             \log \left(\frac{\Lambda_0 }{ \omega + \mu} \right)
\label{s0}
\end{equation}
where $\mu $ measures the deviation of the Fermi level with respect to the Van 
Hove singularity and $\Lambda_0 $ is a high-energy cutoff. The particle-hole 
susceptibility at momentum-transfer ${\bf Q}$ connecting two inequivalent 
saddle points also diverges as 
\begin{equation}
\chi_{\rm ph} ({\bf Q}, \omega ) \approx \frac{1}{2 \sqrt{3} \pi^2} 
 \frac{c'}{\alpha + \beta}  \log \left(\frac{\Lambda_0 }{ \omega + \mu} \right)
\label{sQ}
\end{equation}
with a prefactor given in the case of $\alpha >  3\beta > 0$ by
\begin{equation}
c' =  \log \left(\frac{1+\sqrt{\beta /3\alpha }}{1-\sqrt{\beta /3\alpha }}\right)
   + \log \left(\frac{1+\sqrt{3\beta /\alpha }}{1-\sqrt{3\beta /\alpha }}\right)
\label{pref}
\end{equation}
The singularity in the expression (\ref{pref}) reflects the instance of perfect 
nesting of the Fermi line, which corresponds in the honeycomb lattice to 
$\alpha = 3\beta $ or $\alpha = \beta /3 $. Finally, the particle-particle 
susceptibility $\chi_{\rm pp} ({\bf q}, \omega )$ at vanishing total momentum 
${\bf q}$ of the pair of electrons diverges as
\begin{equation}
\chi_{\rm pp} ({\bf 0}, \omega ) \approx \frac{1}{4 \pi^2} 
     \frac{1}{\sqrt{\alpha \beta }}  
             \log^2 \left(\frac{\Lambda_0 }{ \omega + \mu} \right)
\label{cp0}
\end{equation}

The large growth of the susceptibilities as $\mu \rightarrow 0$ leads to the 
divergence of several response functions as one approaches the Van Hove 
singularity. The strongest divergence among them dictates the character of the 
dominant electronic instability in the system. In this respect, there are two
different scenarios, depending on the relative strength of the particle-hole
susceptibilities at momentum-transfer ${\bf 0}$ and ${\bf Q}$. When
$\chi_{\rm ph} ({\bf Q}, \omega )  >  \chi_{\rm ph} ({\bf 0}, \omega )$, the 
tendency towards a spin-density-wave instability at momentum ${\bf Q}$
prevails over a ferromagnetic instability. In that case, however, a 
superconducting instability may be also viable, due to the anisotropy created 
in the effective $e$-$e$ interaction by the large momentum-dependent
screening near the Van Hove singularity. This is actually the possibility that 
was studied in Ref. \cite{prbkl}. In those circumstances, it was shown  
that the dominant superconducting instability takes place in a $d$-wave 
channel, which corresponds to the degenerate representation 
$\{\cos (2m\theta ) , \sin (2m\theta ) \}$ ($m$ integer, $2m$ not a multiple of 3)
of the point symmetry group $C_{6v}$.

The other scenario corresponds to the case in which  
$\chi_{\rm ph} ({\bf Q}, \omega )  <  \chi_{\rm ph} ({\bf 0}, \omega )$. A 
ferromagnetic instability may arise then in the system, as illustrated in the
analysis of the square lattice by renormalization group methods near the Van 
Hove singularity\cite{ferro} and supported by Monte Carlo simulations\cite{sor}. 
As mentioned before, the case of largest strength of the particle-hole susceptibility 
$\chi_{\rm ph} ({\bf 0}, \omega )$ is in general the relevant instance for 
graphene monolayer and bilayer systems. We will see  in what follows 
that a superconducting instability is also possible in this case, but with an 
order parameter which has preferentially $f$-wave symmetry.

An important point regarding the many-body theory of electrons near a Van Hove
singularity is that all the momentum dependence of the interaction potential 
is irrelevant when scaling the theory towards the low-energy limit. This is 
the reflection of the intense screening from the divergent density of states,
that reduces the effective interaction at low energies to a purely local 
component in real space. This can be seen from inspection of the effective action, 
written in terms of creation (annihilation) operators 
$\psi^\dagger_{\sigma} ({\bf k},t)$ ($\psi_{\sigma} ({\bf k},t)$) 
for electrons with spin $\sigma = \uparrow, \downarrow $ as
\begin{eqnarray}
S  & = & \int dt \: d^2 k \sum_{\sigma} \left(  
    \psi^{\dagger}_{\sigma}({\bf k}, t ) \: i\partial_t \psi_{\sigma}({\bf k}, t )
- \varepsilon ({\bf k}) \; \psi^{\dagger}_{\sigma}({\bf k}, t ) 
   \psi_{\sigma}({\bf k}, t ) \right)                             \nonumber   \\
  &   &  - \frac{1}{2} \int dt \: d^2 k \; 
  \sum_{\sigma , \sigma' } \rho_{\sigma } ({\bf k}, t)  
         \: v_{\sigma \sigma' }  ( {\bf k}) \:  \rho_{\sigma' } (-{\bf k}, t)
\label{actk}
\end{eqnarray}
with the Fourier transform $\rho_{\sigma }({\bf k}, t)$ of the electron density 
given by
\begin{equation}
\rho_{\sigma} ({\bf q}, t )  =    \int  d^2 k \: 
  \psi^\dagger_{\sigma} ({\bf k}+{\bf q}, t)
    \psi_{\sigma} ({\bf k}, t)   
\end{equation}

Thus, one can check that the action (\ref{actk}) is invariant under 
the scale transformation
\begin{equation}
t' = \frac{1}{s} t  \;\; , \;\;  {\bf k}' = \sqrt{s} {\bf k}  \;\; , \;\; 
           \psi' = \frac{1}{\sqrt{s}} \psi 
\label{scale}
\end{equation}
provided one keeps only the zeroth order term in the expansion of 
$v_{\sigma \sigma' }  ( {\bf k})$ in powers of the momentum. Higher orders in 
the power series of the potential would be affected by powers of $1/ \sqrt{s}$, 
with the result that they would be increasingly suppressed as 
$s \rightarrow \infty $. This is the limit in which the theory is scaled down 
to low energies, meaning that one can just remain with the constant term 
$v_{\perp } \equiv  v_{\uparrow \downarrow }  ( {\bf 0})$ for the sake of 
studying the low-energy instabilities of the electron system near the Van Hove
singularity.

\subsection{Ferromagnetic instability}

Focusing on the case where the largest electron-hole susceptibility is given by 
$\chi_{\rm ph} ({\bf 0}, \omega )$, we can sum up in the framework of the RPA
the most divergent contributions to the charge and spin response functions.
These are given by correlators of the electron densities 
$\rho_{\uparrow} ({\bf q},\omega)$ and $\rho_{\downarrow} ({\bf q},\omega)$,
that we define now with more generality
from electron creation (annihilation) operators 
$\psi^\dagger_{j\sigma} ({\bf k},\omega)$ ($\psi_{j\sigma} ({\bf k},\omega)$) 
for a number of independent saddle points $j = 1, \ldots N$,
\begin{equation}
\rho_{\sigma} ({\bf q},\omega_q )  =    \sum_{j,j'}
   \int  d^2 k  \: d \omega_k  \:
  \psi^\dagger_{j\sigma} ({\bf k}+{\bf q},\omega_k+\omega_q)
    \psi_{j'\sigma} ({\bf k},\omega_k)   
\end{equation}
The response functions for charge and spin are given respectively by
\begin{eqnarray}
R_c ({\bf q},\omega )   & = & 
     \langle    \left(  \rho_{\uparrow} ({\bf q},\omega ) 
               +   \rho_{\downarrow} ({\bf q},\omega )   \right)  
                 \;   \left(  \rho_{\uparrow} (-{\bf q},-\omega ) 
             +   \rho_{\downarrow} (-{\bf q},-\omega )   \right)  
                                                             \rangle       \\
R_s ({\bf q},\omega )   & = & 
        \langle      \left(  \rho_{\uparrow} ({\bf q},\omega ) 
                 -   \rho_{\downarrow} ({\bf q},\omega )   \right)  
                 \;   \left(  \rho_{\uparrow} (-{\bf q},-\omega ) 
           -  \rho_{\downarrow} (-{\bf q},-\omega )   \right) 
                                                             \rangle   
\end{eqnarray}  

\begin{figure}
\begin{center}
\mbox{\epsfxsize 14cm \epsfbox{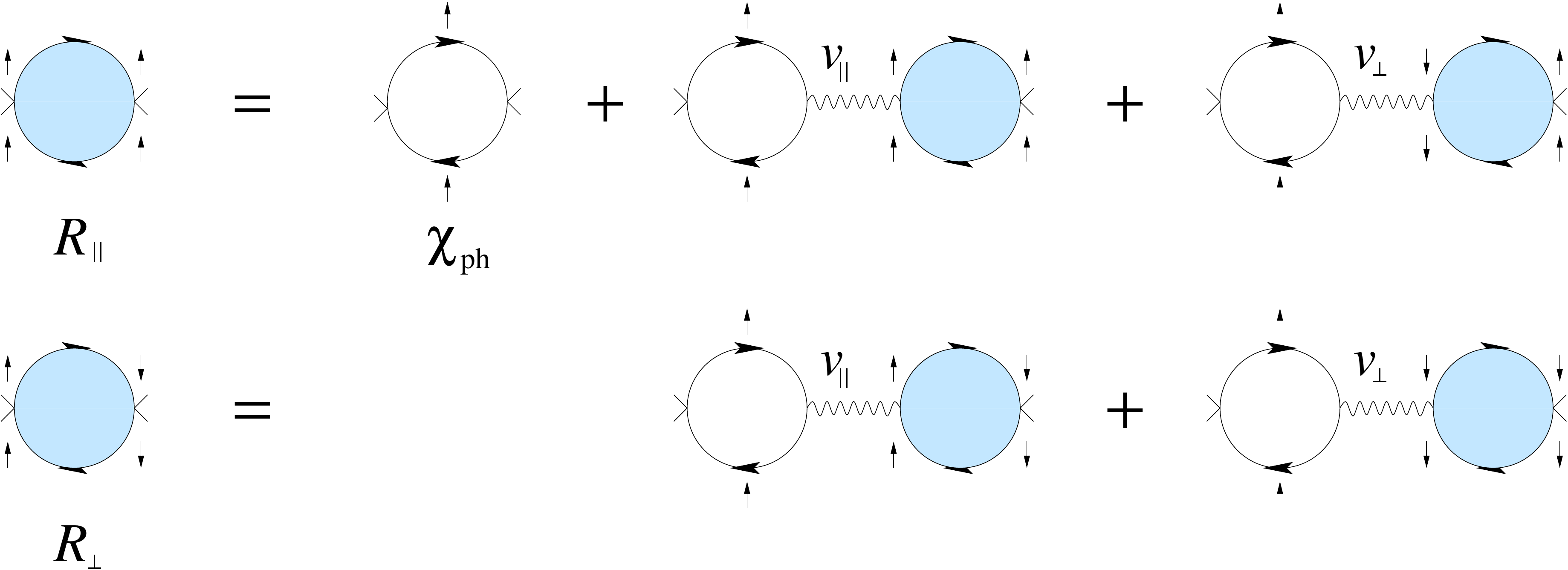}}
\end{center}
\caption{Self-consistent diagrammatic equations for the response functions 
$R_\parallel $ and $R_\perp $ in the RPA.}
\label{diag}
\end{figure}

In general, the $e$-$e$ interaction may be mediated by 
a potential $v_\parallel({\bf q}) $ between electrons with parallel spin and 
$v_\perp({\bf q}) $ for electrons with opposite spin projections. In the RPA, 
the response functions $R_\parallel = (R_c + R_s)/2 $ and 
$R_\perp = (R_c - R_s)/2 $ must obey the self-consistent equations 
represented diagrammatically in Fig. \ref{diag}. We have then
\begin{eqnarray}
R_\parallel ({\bf 0},\omega )  &  =  &   2 N \chi_{\rm ph} ({\bf 0},\omega ) 
 - N \chi_{\rm ph} ({\bf 0},\omega ) \: v_\parallel({\bf 0}) \:  R_\parallel ({\bf 0},\omega ) 
 - N \chi_{\rm ph} ({\bf 0},\omega ) \: v_\perp({\bf 0}) \:  R_\perp ({\bf 0},\omega )  
                                                               \label{rpa}        \\
R_\perp ({\bf 0},\omega )  &  =  &   
 - N \chi_{\rm ph} ({\bf 0},\omega ) \: v_\perp({\bf 0}) \: R_\parallel ({\bf 0},\omega ) 
 - N \chi_{\rm ph} ({\bf 0},\omega ) \: v_\parallel({\bf 0}) \: R_\perp ({\bf 0},\omega )
\label{rpe} 
\end{eqnarray}
Solving the linear system (\ref{rpa})-(\ref{rpe}), we find
\begin{eqnarray}
R_\parallel ({\bf 0},\omega )  & = &  
  \frac{  2 N \chi_{\rm ph} ({\bf 0},\omega ) \: (1 + N v_\parallel({\bf 0}) \chi_{\rm ph} ({\bf 0},\omega ))}
 {(1 + N v_\parallel({\bf 0}) \chi_{\rm ph} ({\bf 0},\omega ))^2 -
                     (N v_\perp({\bf 0}) \chi_{\rm ph} ({\bf 0},\omega ))^2}   \\
R_\perp ({\bf 0},\omega )  & = &  
  \frac{ -2 N^2 v_\perp({\bf 0}) \: (\chi_{\rm ph} ({\bf 0},\omega ))^2}
 {(1 + N v_\parallel({\bf 0}) \chi_{\rm ph} ({\bf 0},\omega ))^2 -
                     (N v_\perp({\bf 0}) \chi_{\rm ph} ({\bf 0},\omega ))^2} 
\end{eqnarray}

We can go back now to the charge and spin response functions, obtaining the result
\begin{eqnarray}
R_c ({\bf 0},\omega )  & = &  \frac{  2 N \chi_{\rm ph} ({\bf 0},\omega )}
  {1 + N (v_\parallel({\bf 0}) + v_\perp({\bf 0})) \: \chi_{\rm ph} ({\bf 0},\omega )}  
                                                       \label{rc}        \\
R_s ({\bf 0},\omega )  & = &  \frac{  2 N \chi_{\rm ph} ({\bf 0},\omega )}
  {1 + N (v_\parallel({\bf 0}) - v_\perp({\bf 0})) \: \chi_{\rm ph} ({\bf 0},\omega )} 
                                                        \label{rs}
\end{eqnarray}
We find then the origin of the spin instability when 
$v_\parallel({\bf 0}) < v_\perp({\bf 0})$. This is the natural situation when 
the screening effects are so strong that the $e$-$e$ interaction is reduced to
a purely on-site repulsion. In the context of electrons interacting near a Van
Hove singularity, we have already seen that $v_\perp({\bf 0})$ is the only 
component of the interaction which is not irrelevant in the low-energy limit. 
This explains why the tendency towards uniform spin order is 
a natural instability in systems with a large density of states for which the 
interaction can be modeled by a local Coulomb repulsion.

A remarkable feature is that the ferromagnetic instability can be always
reached, no matter the strength of the interaction, provided one can place the
Fermi level arbitrarily close to the Van Hove singularity. Taking then the 
expression of the particle-hole susceptibility in (\ref{s0}) and assuming a 
local interaction with constant potential $v_\perp({\bf p}) $, the instability 
is found at the critical energy
\begin{equation}
\omega_c =  \Lambda_0  \exp \left(- \frac{4\pi^2 \sqrt{\alpha \beta}}
                      {N v_\perp}    \right)
\label{wc}
\end{equation}
We have anyhow to bear in mind that a very low value of $\omega_c$ may 
represent the unfeasibility to observe in practice any instability, if 
that energy is below the resolution with which one can approach experimentally
the singularity. This also includes the possible effect of disorder, that may
attenuate the divergence of the density of states below a certain energy scale 
as we will discuss later.

Regarding the expressions (\ref{rc})-(\ref{rs}), we point out that it is 
reassuring their coincidence with the results obtained from a scaling analysis 
of the Van Hove singularity\cite{ferro}. That is, the present results can be 
also interpreted as the lowest order of a renormalization group approach to the 
singularity. This level of approximation still misses relevant effects,
some of them reinforcing the divergent density of states (renormalization of 
the saddle point dispersion) and others tending to weaken the electronic 
correlations (renormalization of the quasiparticle peak)\cite{np}.
Nevertheless, Eq. (\ref{wc}) can be used at least to estimate the order of 
magnitude of the critical energy scale. Full nonperturbative studies of the 
low-energy instabilities arising near a Van Hove singularity (as those carried 
out by means of Monte Carlo simulations in the square lattice\cite{sor}) have 
also certified the existence of the ferromagnetic phase, in regions
of the phase diagram where the condition
$\chi_{\rm ph} ({\bf Q}, \omega )  <  \chi_{\rm ph} ({\bf 0}, \omega )$ is 
satisfied.

\subsection{Pairing instability}

A pairing instability is also possible when the Fermi level is close to the
Van Hove singularity, as the strong screening effects make the effective 
$e$-$e$ interaction quite anisotropic as a function of the momenta of the 
electrons. The tendency towards superconducting order requires however the 
development of an attractive interaction in any of the channels for the 
different representations of the point symmetry group. When this happens, 
the pairing instability can be greatly enhanced due to the divergent density 
of states, already reflected in the particle-particle susceptibility given by 
Eq. (\ref{cp0}). 

The pairing instability can be studied by looking at the behavior of the 
so-called BCS vertex, that is the four-fermion interaction vertex for vanishing 
total momentum and spin of the incoming electrons. Such a function can be 
parametrized in terms of the angles $\theta $ and $\theta'$ of the respective
momenta of the spin-up incoming and outgoing electrons. Henceforth we will 
therefore denote the BCS vertex by $V(\theta, \theta'; \omega)$,
$\omega $ being the energy of the pair of electrons. 

\begin{figure}
\begin{center}
\mbox{\epsfxsize 10cm \epsfbox{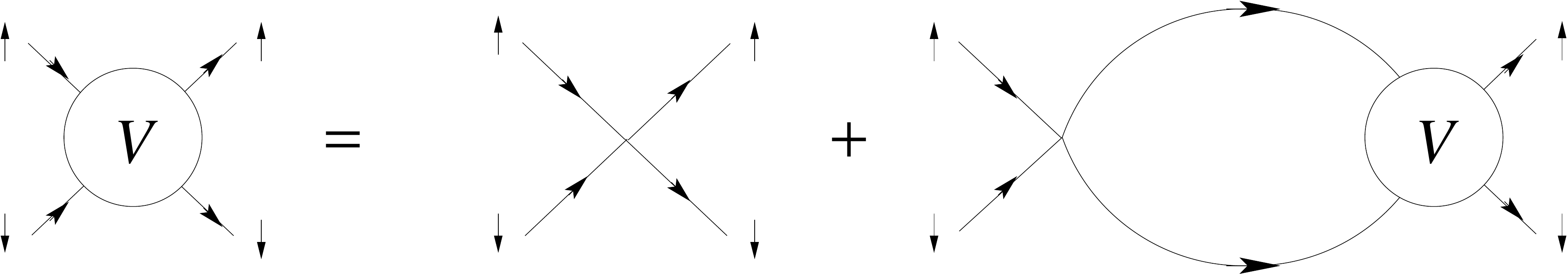}}
\end{center}
\caption{Self-consistent diagrammatic equation for the BCS vertex $V$ in the ladder
approximation.}
\label{cooper}
\end{figure}

Adopting a similar methodology as for the ferromagnetic instability, we will 
seek to sum up the most divergent contributions to the BCS vertex
function. These can be encoded in the diagrammatic equation represented in 
Fig \ref{cooper}, where the second term at the right-hand-side accounts for the
divergence of the particle-particle susceptibility. The particle-particle loop
involves an integration in momentum space, that can be parametrized in terms of 
the differential elements $d k_\parallel $ and $d k_\perp $, longitudinal and 
normal respectively to the lines of constant energy. Alternatively, one can 
pass to integration variables defined by the energy $\varepsilon $ of the 
contour lines and the angle $\theta $ along them. Thus, we end up with the 
self-consistent equation
\begin{equation}
V(\theta, \theta'; \omega) = V_0 (\theta, \theta') - 
  \frac{1}{(2\pi )^2} \int_0^{\Lambda } d \varepsilon \int_0^{2\pi } d \theta'' 
  \frac{\partial k_\perp }{\partial \varepsilon}  \frac{\partial k_\parallel }{\partial \theta''} 
  V_0 (\theta, \theta'')   \frac{1}{\varepsilon  - \frac{\omega }{2}}  
         V(\theta'', \theta'; \omega)
\end{equation}

At this point, one can make contact with the more powerful scaling approach by 
differentiating with respect to the high-energy cuttof $\Lambda $ and applying 
the self-consistency to the right-hand-side of the equation. Then we get
\begin{equation}
\frac{\partial V(\theta, \theta'; \omega)}{\partial \Lambda } = 
 - \frac{1}{(2\pi )^2}  \int_0^{2\pi } d \theta'' 
  \frac{\partial k_\perp }{\partial \varepsilon}  \frac{\partial k_\parallel }{\partial \theta''} 
  V (\theta, \theta''; \omega)   \frac{1}{\Lambda}  
         V(\theta'', \theta'; \omega)
\label{rge}
\end{equation}      
This is precisely the equation that one obtains in a renormalization group 
approach to the pairing instability and, in the present context, it has the 
advantage of allowing a proper consideration of the effect of the divergent
density of states upon scaling in the low-energy limit. The density of states
can be actually expressed as 
\begin{equation}
n(\Lambda ) =  \frac{1}{(2\pi )^2} \int_0^{2\pi } d \theta'' 
  \frac{\partial k_\perp }{\partial \varepsilon}  \frac{\partial k_\parallel }{\partial \theta''}
\end{equation}
where the integral is carried out along a contour line of energy $\Lambda $. 
Then we can write Eq. (\ref{rge}) is simpler form by passing to the angular 
variable
\begin{equation}
\phi (\theta ) = \frac{1}{2\pi n(\Lambda )} \int_0^{\theta }  d\theta'' 
   \frac{\partial k_\perp }{\partial \varepsilon}  \frac{\partial k_\parallel }{\partial \theta''}
\end{equation}
In terms of the new function $\widetilde{V} (\phi, \phi'; \omega) = V (\theta, \theta'; \omega)$, 
the scaling equation reads
\begin{equation}
\Lambda  \frac{\partial \widetilde{V} (\phi, \phi'; \omega)}{\partial \Lambda } = 
 - \frac{n(\Lambda )}{2\pi } \int_0^{2\pi } d\phi'' \widetilde{V} (\phi, \phi''; \omega) 
      \widetilde{V} (\phi'', \phi'; \omega)
\label{rgn}
\end{equation}

The BCS vertex is a function of the frequency $\omega $ as well as of
the high-energy cutoff $\Lambda $. Under the assumption of scaling, it must be
actually a function of the ratio $\omega /\Lambda $, so the low-energy limit 
$\omega \rightarrow 0$ can be approached by solving Eq. 
(\ref{rgn}) in the limit of large $\Lambda $. In the particular case of 
approximately constant density of states $n$ and angle-independent vertex 
$\widetilde{V}(\omega) $, the solution of the equation leads to a singularity 
for attractive interaction $\widetilde{V} < 0$, with the well-known relation
between the low-energy critical scale $\omega_c$ and the high-energy cutoff
\begin{equation}
\omega_c  \approx  \Lambda_0  \exp \{ -1/\lambda \}
  \;\;\;\;\;\;\;\; ,  \;\;\;\;\;\;\;\; \lambda \equiv  n|\widetilde{V} (\Lambda_0)|
\label{bcs}
\end{equation}
In the proximity of the Van Hove singularity, however, we see that a pairing
instability may be enhanced by the logarithmic divergence of $n(\Lambda )$,
increasing significantly the value of $\omega_c$.

In the case of energy-dependent density of states, we can still resort to an 
approximation that allows to estimate the strength of a pairing instability
by using a relation like (\ref{bcs}), in terms of a set of constant 
effective couplings. For that purpose, we can introduce in Eq. (\ref{rge})
the change of variables
\begin{equation}
\widehat{V} (\theta, \theta'; \omega)  =
\sqrt{ \frac{1}{2\pi } \frac{\partial k_\perp (\theta)}{\partial \varepsilon}
  \frac{\partial k_\parallel (\theta)}{\partial \theta }  }
\sqrt{ \frac{1}{2\pi } \frac{\partial k_\perp (\theta')}{\partial \varepsilon}
  \frac{\partial k_\parallel (\theta')}{\partial \theta' }  }
    V (\theta, \theta'; \omega)
\label{redef}
\end{equation} 
In terms of the redefined vertex, the scaling equation can be then approximated 
by
\begin{equation}
\Lambda \frac{\partial \widehat{V}(\theta, \theta'; \omega)}{\partial \Lambda } 
 =  - \frac{1}{2\pi }  \int_0^{2\pi } d \theta''  
 \widehat{V} (\theta, \theta''; \omega)  \widehat{V}(\theta'', \theta'; \omega)
\label{scaling}
\end{equation} 
The vertex (\ref{redef}) is actually the starting point of usual 
analyses of the pairing instabilities near a Van Hove 
singularity\cite{slh,lub}, where the couplings computed for the effective
attraction can be interpreted in the framework of the standard BCS
theory.

The integration of Eq. (\ref{scaling}) can be facilitated by expanding 
the vertex $\widehat{V} (\theta, \theta'; \omega)$ in terms of the modes 
$\Psi_m^{(\gamma )} (\theta)$ for the different representations $\gamma $ of the 
point symmetry group,
\begin{equation}
\widehat{V} (\theta, \theta'; \omega) = \sum_{\gamma, m, n} V_{m,n}^{(\gamma )}
       \Psi_m^{(\gamma )} (\theta)   \Psi_n^{(\gamma )} (\theta')
\label{dec}
\end{equation}
We arrive then at a set of equations for each representation $\gamma $
\begin{equation}
\Lambda  \frac{\partial V_{m,n}^{(\gamma )}}{\partial \Lambda } =
 -  \sum_{s}  V_{m,s}^{(\gamma )}  V_{s,n}^{(\gamma )}
\label{rg}
\end{equation}
It becomes clear that, for 
positive initial values of $V_{m,n}^{(\gamma )}$, the couplings fade
away in the low-energy regime approached as $\Lambda \rightarrow \infty $. 
In the present case of highly anisotropic
screening, some of the channels may start however with an attractive effective 
interaction. Then, as observed from Eq. (\ref{rg}), this will be enough to 
trigger a pairing instability at a low-energy critical scale like that in Eq.
(\ref{bcs}). 

A sensible way of computing initial values for $V_{m,n}^{(\gamma )}$ 
is to start with a dressed vertex accounting for the effects of the 
electron-hole polarization not included in the sum of Fig. \ref{cooper}. One
can indeed perform the sum of RPA and ladder contributions obtained from 
iteration of the electron-hole scattering, as proposed in Ref. \cite{slh}. 
Assuming as in the previous section a local interaction with constant 
potential $v_\perp $ in momentum space, the initial value of the BCS 
vertex is given in this approximation by\cite{slh} 
\begin{equation}
 \widehat{V}_0 (\theta, \theta') =  F(\theta )  F(\theta' ) \left( v_\perp + 
  \frac{v_\perp^2 \chi_{\rm ph}(\mathbf{k}+\mathbf{k}^\prime)}
         {1 - v_\perp \chi_{\rm ph}(\mathbf{k}+\mathbf{k}^\prime)}
 + \frac{v_\perp^3 \chi_{\rm ph}^2(\mathbf{k}-\mathbf{k}^\prime)}
       {1 - v_\perp^2 \chi_{\rm ph}^2(\mathbf{k}-\mathbf{k}^\prime)}  \right)
\label{init}
\end{equation} 
with 
\begin{equation}
F(\theta ) =  
 \sqrt{ \frac{1}{2\pi } \frac{\partial k_\perp (\theta)}{\partial \varepsilon}
  \frac{\partial k_\parallel (\theta)}{\partial \theta }  }
\end{equation}
and $\mathbf{k}, \mathbf{k}^\prime$ being the respective momenta at
angles $\theta, \theta'$ over the energy contour line.
The singularity in the fractions of Eq. (\ref{init}) corresponds at vanishing 
momentum to the ferromagnetic instability discussed in the previous section. 
As already mentioned, the mode expansion of (\ref{init}) may lead however to a 
negative  coupling in some of the channels, making the pairing instability to
prevail at a higher critical scale, as we will see in the next section.

\section{Van Hove singularities in monolayer and twisted bilayer graphene}

In all the relevant instances where a Van Hove singularity arises in 
graphene-based systems, the electron-hole susceptibility appears to reach its
maximum peak at vanishing momentum transfer. This narrows down the 
possible electronic instabilities to either a tendency towards ferromagnetism 
or towards Cooper pairing with unconventional (preferentially $f$-wave) 
order parameter. The case of monolayer graphene with large electron doping is 
an example of the latter instance, as a result of the strong modulation of the 
interaction around the extended saddle points in the electronic dispersion. 
The tendency to magnetic order turns out to be dominant instead in the twisted 
graphene bilayers, given the lower degree of symmetry but larger 
number of saddle points in the dispersion.

\subsection{Monolayer graphene}

\begin{figure}
\begin{center}
\mbox{\epsfxsize 6.0cm \epsfbox{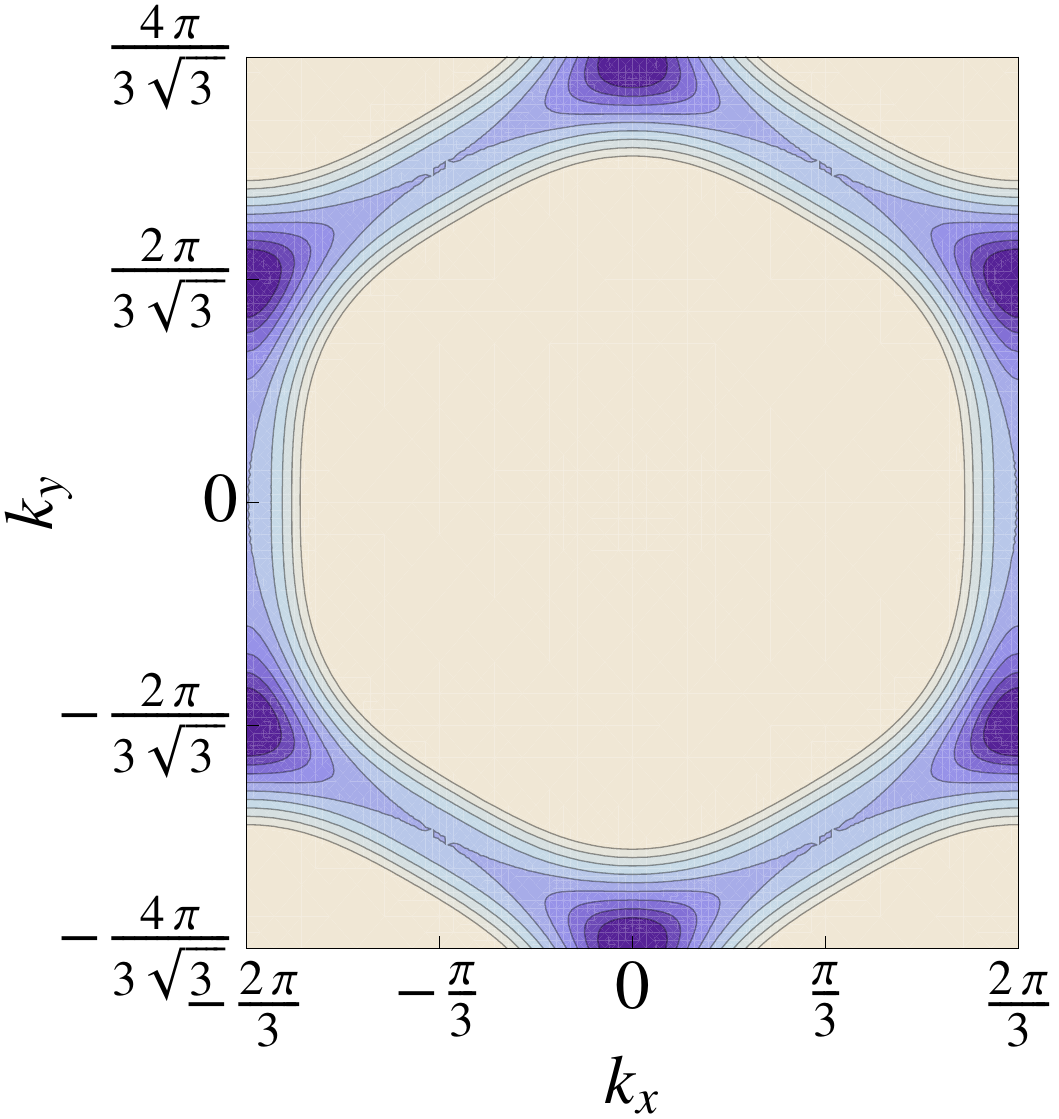}}
\end{center}
\caption{Plot of energy contour lines around the saddle points in the 
conduction band of graphene, obtained from a tight-binding model with first, 
second, and third-neighbor hopping parameters (momenta $k_x$ and $k_y$ are 
measured in units of the inverse of the C-C distance).}
\label{vhmon}
\end{figure}

The electronic dispersion near the Van Hove singularity in the conduction band
of graphene has been mapped in the ARPES experiments reported in Ref. 
\cite{prl}. The saddle point dispersion around the M point of the Brillouin 
zone shows an extended character that cannot be accounted for by conventional
approaches like the LDA or GW approximations, reflecting that it  arises
as an effect of strong electronic correlation. One can anyhow fit empirically
the electronic dispersion by writing down a tight-binding hamiltonian for 
electron creation (annihilation) operators $c_i^\dagger $ ($c_i$) in the 
honeycomb lattice
\begin{equation}
H =  - t \sum_{i,j} c_i^\dagger c_j - t' \sum_{i,j} c_i^\dagger c_j 
                  - t'' \sum_{i,j} c_i^\dagger c_j
\label{tb}
\end{equation}
where the sums account respectively for hopping between first, second and 
third-neighbor carbon atoms. The hopping parameters giving the best fit 
to the experimental results can be found in Ref. \cite{prl}. A plot of the 
energy contour lines obtained from the diagonalization of (\ref{tb}) near 
the Van Hove singularity in the conduction band is shown in Fig. 
\ref{vhmon}.

\begin{figure}[t]
\begin{center}
\mbox{\epsfxsize 6.0cm \epsfbox{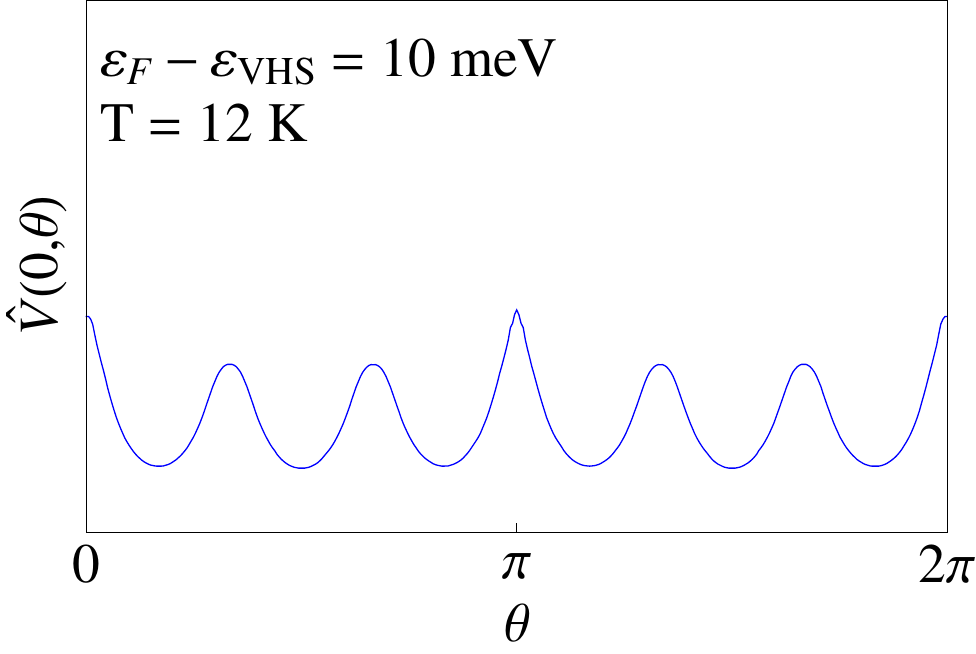}}
\hspace{1cm}
\mbox{\epsfxsize 6.0cm \epsfbox{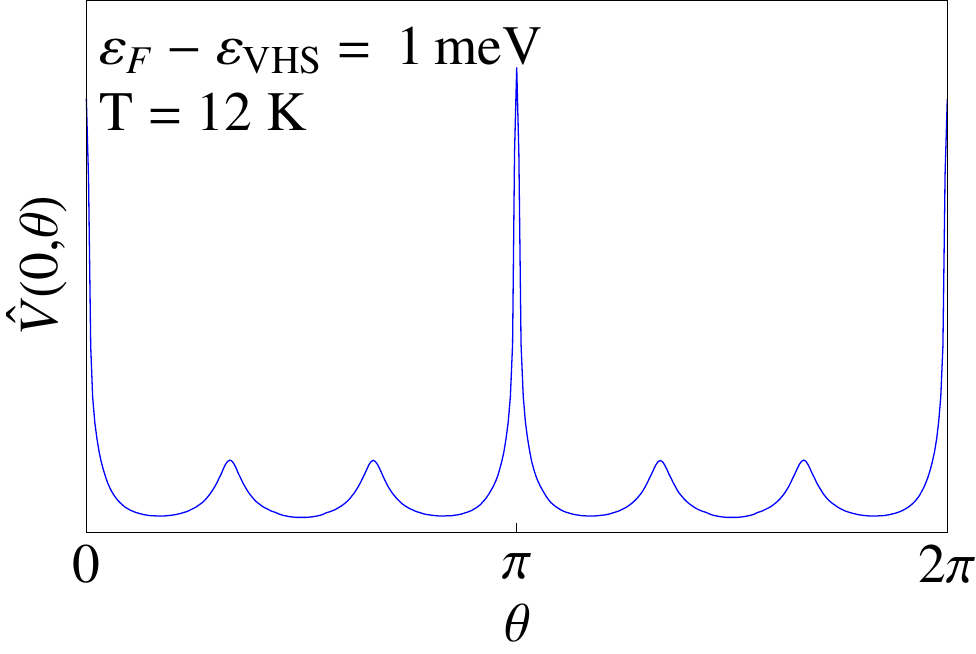}}
\end{center}
\hspace{0.2cm}  (a)  \hspace{6.6cm}  (b)
\caption{Modulation of the BCS vertex as the angle $\theta $ makes a 
complete turn along the Fermi line, for different shifts of the Fermi energy 
$\varepsilon_F$ with respect to the level of the Van Hove singularity 
$\varepsilon_{\rm VHS}$ and temperature $T = 12$ K.}
\label{vmod}
\end{figure}

The extended character of the saddle points seen in Fig. \ref{vhmon} leads
to a marked enhancement of the electron-hole susceptibility at vanishing 
momentum transfer. Moreover, the BCS vertex develops a strong modulation along 
the Fermi line passing near the saddle points, as observed in Fig. \ref{vmod}. 
The oscillations are translated to the initial condition (\ref{init}) for the 
vertex, in such a way that the periodic behavior along the Fermi line is 
dominated by the first harmonics:
\begin{eqnarray}
\widehat{V}_0 (\theta, \theta)  & = &  c_0 + c_2 \cos (2\theta ) +
     c_4 \cos (4\theta )  +   c_6 \cos (6\theta )  +   \ldots     \label{exp1}  \\
\widehat{V}_0 (0, \theta)  & = &   c_0' + c_2' \cos (2\theta ) +
     c_4' \cos (4\theta )  +   c_6' \cos (6\theta )  +   \ldots            \\
\widehat{V}_0 (\theta, -\theta)  & = &   c_0'' + c_2'' \cos (2\theta ) +
     c_4'' \cos (4\theta )  +   c_6'' \cos (6\theta )  +   \ldots
\label{exp3}
\end{eqnarray}

The expansions (\ref{exp1})-(\ref{exp3}) match well with the periodicity of
the modes for the irreducible representations of the $C_{6v}$ symmetry group. 
Four of them are one-dimensional, with respective sets of basis functions given
by $\{ \cos (6n\theta) \}$, $\{ \sin (6n\theta) \}$, $\{ \cos ((6n+3)\theta) \}$ and
$\{ \sin ((6n+3)\theta) \}$ ($n$ being always an integer). The remaining two 
representations are two-dimensional, corresponding to the sets 
$\{ \cos (m \theta) , \sin (m \theta) \}$, with the integer $m$ running over 
all values that are not multiple of 3 and which are odd for one of the 
representations and even for the other. The dominant terms in 
(\ref{exp1})-(\ref{exp3}) can be accounted for by approximating the BCS 
vertex with the first modes of the irreducible representations:
\begin{eqnarray}
 \widehat{V} (\theta, \theta'; \omega)   & = &    V_{0,0} 
 + 2 V_{2,2} [ \cos (2\theta ) \cos (2\theta') + \sin (2\theta ) \sin (2\theta') ]   
                                                                \nonumber    \\
 & & +  2 V_{2,4} [ \cos (2\theta ) \cos (4\theta') - \sin (2\theta ) \sin (4\theta') + 
             \theta \leftrightarrow \theta'  ]          \nonumber     \\
 & & +  2 V_{3,3} \sin (3\theta ) \sin (3\theta')               
       +  2 V_{3,3}'  \cos (3\theta ) \cos (3\theta')               \nonumber    \\
 & &         + \sqrt{2} V_{0,6} [ \cos (6\theta ) + \cos (6\theta' ) ]   +   \ldots    
\label{exp}
\end{eqnarray}
Comparing (\ref{exp}) with (\ref{exp1})-(\ref{exp3}), one can draw easily the 
correspondence $c_6 = 4V_{2,4} - V_{3,3} + 2 \sqrt{2} V_{0,6}, 
c_2' = 2 V_{2,2} + 2 V_{2,4}, c_4' = 2 V_{2,4}, c_6' = \sqrt{2} V_{0,6}$, and 
$c_2'' = 4 V_{2,4}, c_4'' = 2 V_{2,2}, c_6'' = V_{3,3} + 2 \sqrt{2} V_{0,6}$.

Using very simple arguments, it is possible to show that at least one of the 
couplings in the expansion (\ref{exp}) must be negative, for the particular 
case of dispersion with the shape depicted in Fig. \ref{vhmon}. We may take
for instance two specific pairs of angles for which the BCS vertex is amplified 
by electron-hole scattering, namely  
\begin{equation}
\widehat{V}  \left(\tfrac{\pi }{6}, \tfrac{\pi }{6} ;\omega \right)  
  \approx  V_{0,0} + 2 V_{2,2}  - 4 V_{2,4}  + 2 V_{3,3} - 2 \sqrt{2} V_{0,6}
\end{equation}
and 
\begin{equation}
\widehat{V} \left(\tfrac{\pi }{2}, -\tfrac{\pi }{2} ;\omega \right)   
  \approx  V_{0,0} + 2 V_{2,2} - 4 V_{2,4}  - 2 V_{3,3} - 2 \sqrt{2} V_{0,6}
\end{equation}  
Looking now at the initial condition given by Eq. (\ref{init}), it becomes 
clear that
\begin{equation}
\widehat{V}_0  \left(\tfrac{\pi }{6}, \tfrac{\pi }{6}  \right) 
  <  \widehat{V}_0 \left(\tfrac{\pi }{2}, -\tfrac{\pi }{2}  \right) 
\label{neg}
\end{equation}
since $\widehat{V}_0 (\pi /2, -\pi /2 )$ is enhanced by the 
term depending on $\chi_{\rm ph}(\mathbf{k}+\mathbf{k}^\prime)$ in (\ref{init})
at $\mathbf{k}^\prime = - \mathbf{k} $, while 
$\widehat{V}_0 (\pi /6, \pi /6 )$ is enhanced by the weaker term that 
depends on $\chi_{\rm ph}(\mathbf{k}-\mathbf{k}^\prime)$ at 
$\mathbf{k}^\prime = \mathbf{k} $.
On the other hand, we have 
\begin{equation}
\widehat{V}  \left(\tfrac{\pi }{6}, \tfrac{\pi }{6} ;\omega \right) 
  -  \widehat{V} \left(\tfrac{\pi }{2}, -\tfrac{\pi }{2} ;\omega \right) 
  \approx  4 V_{3,3}
\label{v33}
\end{equation}
implying that $V_{3,3}$ must be negative.

\begin{figure}[t]
\begin{center}
\mbox{\epsfxsize 6.0cm \epsfbox{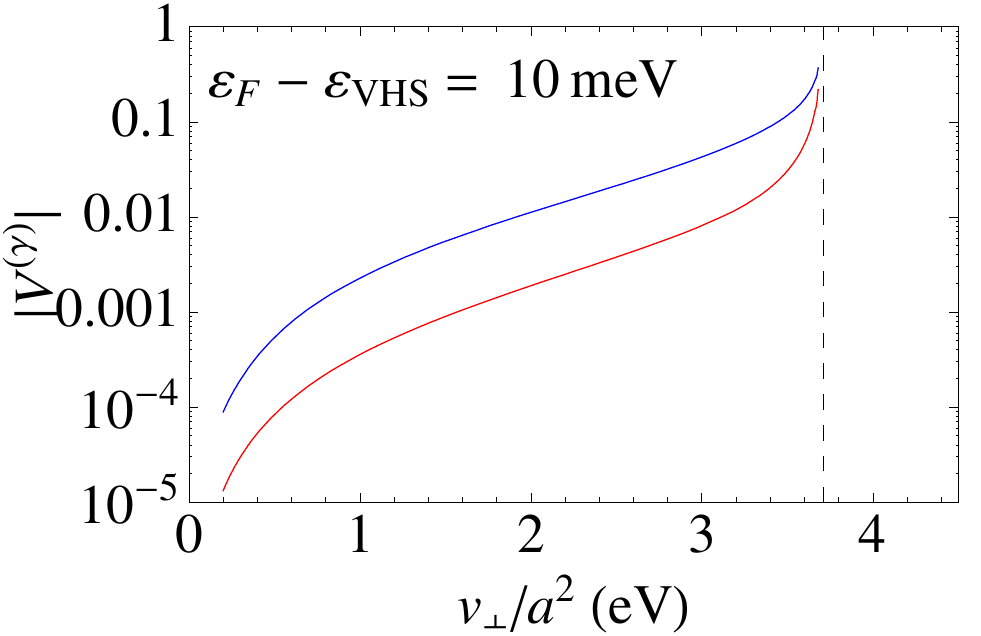}}
\hspace{1cm}
\mbox{\epsfxsize 6.0cm \epsfbox{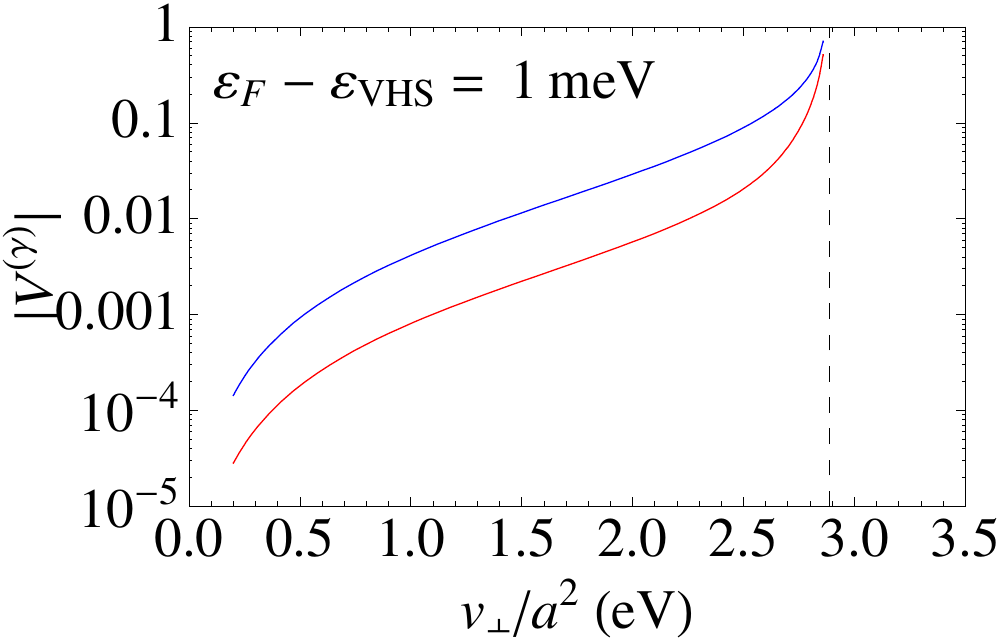}}
\end{center}
\hspace{0.2cm}  (a)  \hspace{6.6cm}  (b)
\caption{Plot of the absolute value of the negative couplings $V^{(\gamma )}$ 
in the channels with $\sin (3\theta )$ symmetry (upper curves) and
$\cos (3\theta )$ symmetry (lower curves) as functions of the ratio of the 
potential $v_\perp$ to the square of the C-C distance $a$. The dashed lines 
represent the location of the singularities in the fractions of the expression
(\ref{init}).}
\label{evo}
\end{figure}

It can be checked that $V_{3,3}$ is actually the dominant negative coupling 
among the terms that appear in the expression (\ref{exp}). We arrive therefore 
at the conclusion that $f$-wave must be the symmetry of the dominant pairing 
instability at weak coupling, when only the first terms are significant in the
expansion of the vertex. This 
is in agreement with the results reported in Ref. \cite{prl} near the Van Hove 
singularity in the conduction band of graphene, where it was found a dominant 
superconducting instability with $f$-wave order parameter over most part of 
the phase diagram (we note that the instability was assigned in that reference 
to the representation with $\cos (3\theta)$ symmetry, consequent with the fact
that the axes were rotated by $\pi/2$ with respect to the present notation). 
The actual values of the negative couplings can be seen for two different 
doping levels in Fig. \ref{evo}, which represents the projections 
$V^{(\gamma )}$ of the vertex (\ref{init}) after its numerical evaluation in a 
grid of $800 \times 800$ points in the Brillouin zone of the honeycomb lattice.

Very close to the Van Hove singularity, there could be still room for a 
different pairing instability with $d$-wave symmetry\cite{prl}, though the 
analysis cannot be based then on a simple picture like that from the expression 
(\ref{exp}). Remaining otherwise with the weak coupling expansion, the scaling 
equation (\ref{rg}) becomes in the sector of the representation 
$\{ \cos (m \theta) , \sin (m \theta) \}$ for $m$ even
\begin{equation}
\Lambda  \frac{\partial }{\partial \Lambda } 
\left(
\begin{array}{cc}
V_{2,2}  &   V_{2,4}    \\
V_{4,2}    &   V_{4,4}
\end{array}         \right)
\approx -
\left(
\begin{array}{cc}
V_{2,2}  &   V_{2,4}    \\
V_{4,2}    &   V_{4,4}
\end{array}         \right)
\left(
\begin{array}{cc}
V_{2,2}  &   V_{2,4}    \\
V_{4,2}    &   V_{4,4}
\end{array}         \right)
\label{mrg}
\end{equation}
This equation can be integrated by passing to the eigenvalues of the 
matrix of couplings, 
\begin{equation}
\lambda_{1,2} = \tfrac{1}{2} (V_{2,2} +  V_{4,4}) 
      \pm \tfrac{1}{2}  \sqrt{(V_{2,2} - V_{4,4})^2 + 4V_{2,4}^2}
\end{equation}
In the present situation where the 
electron-hole polarization is largest at vanishing momentum transfer, both
couplings $V_{2,2}$ and $V_{4,4}$ are positive and comparable in the expansion 
(\ref{exp}). This means that both eigenvalues turn out to be positive, so that 
a $d$-wave instability cannot exist over most part of the phase diagram, away 
from the strong coupling regime in which $v_\perp \chi_{\rm ph}$ is close to 
the poles in the expression (\ref{init}). 
The same consideration can be applied to a $p$-wave instability. 
It is only for doping levels very close 
to the Van Hove singularity, where the simple approximation (\ref{exp}) does 
not apply and many more modes start to contribute significantly to the 
expansion, that other instabilities apart from the mentioned $f$-wave may come 
into play. The precise analysis depends then on the particular shape of the 
dispersion but, as the numerical computation of Ref. \cite{prl} shows, the 
instability with $f$-wave symmetry appears to be dominant even for relatively
low doping levels, until the Fermi energy is tuned within $\sim 0.1$ meV
about the Van Hove singularity.

%where the Hubbard-like interaction $U$ (in eV) is given by the 
%ratio of the coupling $v_{\perp}$ per Angstr\"om square.

At this point, it is interesting to compare the above results with those
obtained in Ref. \cite{prbkl} for the case of largest electron-hole 
scattering at a momentum ${\bf Q}$ connecting the saddle points in the 
electron dispersion. Under the assumption that 
$\chi_{\rm ph} ({\bf Q}, \omega )  >  \chi_{\rm ph} ({\bf 0}, \omega )$,
it was shown there that the couplings $V_{2,2}$ and $V_{2,4}$ are both 
negative, implying the development of a dominant pairing instability always
with $d$-wave symmetry. The prevalence of  $\chi_{\rm ph} ({\bf Q}, \omega )$
requires however a condition of approximate nesting of the Fermi line near 
the saddle points. As we have seen, this is very far from being realized in
the conduction band of graphene, which stresses the role of a largest 
electron-hole scattering at ${\bf q} = 0$ to account for the pairing 
instabilities found in the present case.

In principle, a ferromagnetic instability could also compete with the tendency
to Cooper pairing in the presence of a large susceptibility 
$\chi_{\rm ph} ({\bf 0},\omega )$. At a given temperature and doping level, 
there is a critical interaction $(v_\perp)_c$ at which the ferromagnetic 
instability can take place, determined by the location of the singularity
in the fractions of the expression (\ref{init}). It is 
important to note however that the own negative couplings $V^{(\gamma )}$
derived for the pairing instability diverge as that critical interaction 
strength is approached. This is clearly appreciated in the evolution of the 
couplings shown in Fig. \ref{evo}. That is, the extended character of the 
saddle points makes the singularity in the vertex (\ref{init}) nonintegrable
along the Fermi line. 
%This means that, before the critical interaction 
%$(v_\perp)_c$ at a given temperature $T_{\rm ferro}$ is reached, the critical 
%temperature for the pairing instability is able to exceed that value of 
%$T_{\rm ferro}$ at a lower $v_{\perp } < (v_\perp)_c$. 
This means that, before the temperature $T_{\rm ferro}$ of the singularity for 
a given critical $(v_\perp)_c$ is reached, the critical 
temperature $T_c$ for the pairing instability is able to exceed that value of 
$T_{\rm ferro}$ as the coupling $V^{(\gamma )}$ diverges (in the limit 
$v_\perp  \rightarrow  (v_\perp)_c $) in the expression 
\begin{equation}
k_B T_c  \approx  \Lambda_0  \exp \{ -1/|V^{(\gamma )}| \}
\label{ctemp}
\end{equation}
Consequently, it is 
clear that the pairing instability must prevail over the ferromagnetic 
instability, in this particular case of the Van Hove singularity in the 
conduction band of graphene.

\subsection{Twisted graphene bilayer}

{\em Band structure.---} Twisted graphene bilayers are a class of coupled 
graphene layers in
which there is a relative rotation between the symmetry axes of the two carbon 
sheets. This gives rise to characteristic Moir\'e patterns showing the periodic 
repetition of an alternating stacking with the form of a hexagonal 
superlattice\cite{port}. We are going to deal here with a description of the 
twisted bilayers for relatively large period of the Moir\'e pattern, which will 
allow us to apply a continuum approach to the interactions among the large 
number of atoms in the unit cell of the superlattice. This has in general 
primitive vectors ${\bf L}_\pm = L (\sqrt{3}/2,\pm 1/2)$, where the period $L$ 
is given in terms of the twist angle $\theta $ by the relation 
$L = a_0/[2 \sin (\theta /2) ]$ ($a_0$ being the lattice constant of graphene). 
Starting from perfect Bernal stacking at $\theta = \pi/3 $, the values of the 
twist angle 
consistent with a commensurate superlattice are quantized, giving rise to a 
sequence of increasing periods
\begin{equation} 
L_n = \sqrt{3 n^2 + 3 n + 1} \: a_0
\label{seq}
\end{equation}
for integer $n \geq 0$ \cite{port}. 

\begin{figure}[t]
\begin{center}
\mbox{\epsfxsize 5.0cm \epsfbox{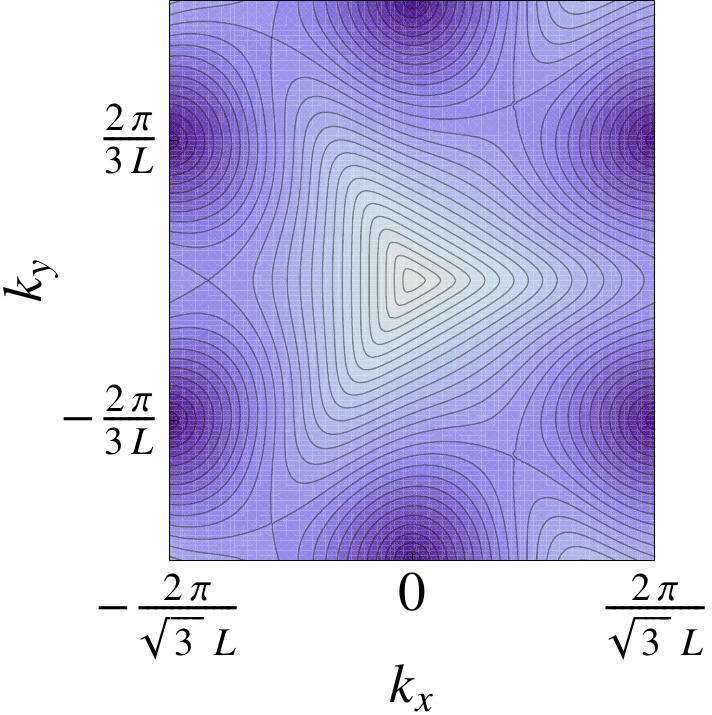}}
\hspace{1cm}
\mbox{\epsfxsize 5.0cm \epsfbox{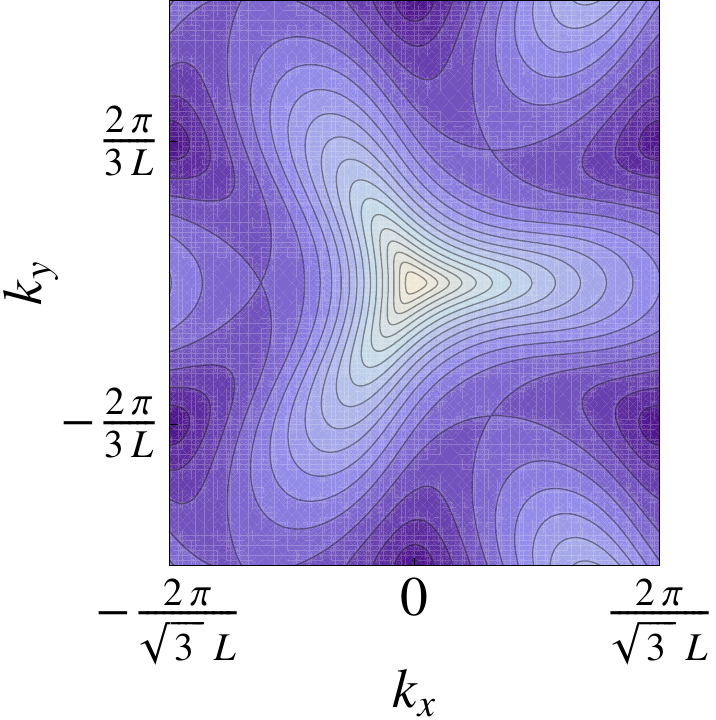}}
\hspace{1cm}
\mbox{\epsfxsize 5.0cm \epsfbox{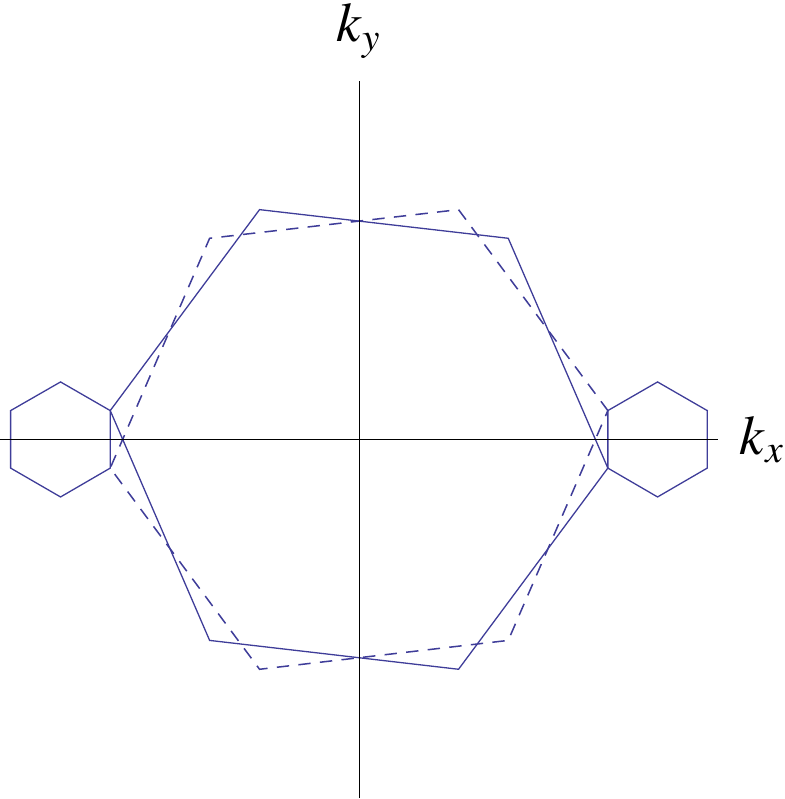}}
\end{center}
\hspace{0cm}  (a)  \hspace{5.6cm}  (b)  \hspace{5.1cm}  (c)
\caption{(a) and (b): Plot of energy contour lines showing the saddle points 
in the lowest-energy subband of twisted graphene bilayers corresponding 
respectively to $n = 10$ and 22 in the sequence (\ref{seq}). (c) Scheme 
showing the relative rotation of the Brillouin zones of the layers (large 
hexagons) and the relative position of the two regions (small hexagons) making 
up the Brillouin zone of the twisted bilayer.}
\label{vhtwist}
\end{figure}

The band structure of the twisted bilayer can be obtained by realizing first 
that there must be a relative twist between the Brillouin zones of the two 
carbon layers, given by the angle of rotation $\theta $. This leads to a 
mismatch in the position of the respective $K$ points which, if originally 
placed at ${\bf K} = (4\pi /3 a_0, 0)$, can be taken as shifted in opposite 
directions by $\pm \Delta {\bf K}/2 = (0, \pm |{\bf K}| \sin (\theta /2))$ 
(see Fig. \ref{vhtwist}(c)). Next, one has 
to account for the hybridization of states in the $\pi $ bands of the two 
layers, which can undergo scattering with a momentum transfer dictated by the
periodic structure of the Moir\'e superlattice. This leads to folding of the 
bands of the twisted bilayer into an hexagonal Brillouin zone, with reciprocal
vectors ${\bf Q}_{1,2} = (2\pi /L) (\pm 1/\sqrt{3}, 1)$ \cite{port}.

For relatively large values of $L$ compared to $a_0$, the low-energy physics 
of the twisted bilayer can be extracted from the hybridization between states 
around the respective Dirac points of the two layers rotated by a small twist 
$\theta $. In this approach, one can make a microscopic average of the 
tunneling amplitude between the two graphene layers, represented by smooth 
interlayer potentials $V_{AA'}({\bf r}), V_{AB'}({\bf r})$ and 
$V_{BA'}({\bf r})$ accounting for the modulated hopping between sublattices 
$A,B$ of one layer and $A',B'$ of the other layer. In the space of 
four-component spinors $( \psi_A , \psi_B , \psi_{A'} , \psi_{B'} )$ made
of the electronic amplitudes on each sublattice  of the two graphene layers, 
the hamiltonian can be written in the form\cite{prl3}
\begin{eqnarray}
H=
v_F \left(
 \begin{array}{cccc}
0 &  -i\partial_x - \partial_y + i \Delta|{\bf K}|/2 & 
                           V_{AA'}(\mathbf{r}) & V_{AB'}(\mathbf{r}) \\
 -i\partial_x + \partial_y - i \Delta|{\bf K}|/2 & 0 & 
                           V_{BA'}(\mathbf{r}) & V_{AA'}(\mathbf{r}) \\
 V_{AA'}^\star(\mathbf{r}) & V_{BA'}^\star(\mathbf{r}) & 0 & 
                         -i\partial_x - \partial_y - i \Delta|{\bf K}|/2 \\
 V_{AB'}^\star(\mathbf{r}) & V_{AA'}^\star(\mathbf{r}) &  
                        -i\partial_x + \partial_y + i \Delta|{\bf K}|/2 & 0
 \end{array}\right)
\label{H}
\end{eqnarray}
This provides the pertinent construction around a pair of Dirac points shifted
by $\pm \Delta|{\bf K}|/2$ with respect to the original $K$ points of the 
graphene layers. One has anyhow to bear in mind that the complete spectrum must 
be obtained by adding the contribution of a similar hamiltonian representing 
the hybridization of the Dirac cones at the opposite $K$ points, which can be 
obtained by reversing the sign of the $x$ variable and exchanging the two 
layers in Eq. (\ref{H}).  

The lowest-energy subbands of the twisted graphene bilayers can be found by 
diagonalizing the hamiltonian with a sensible representation of the interlayer
potentials, in accordance with the symmetry of the hexagonal superlattice. This
implies in particular the periodicity
$V_{AA'} (\mathbf{r}) = V_{AA'} (\mathbf{r}+\mathbf{L}_+) = 
V_{AA'} (\mathbf{r}+\mathbf{L}_-)$, 
and the relations 
$V_{AB'} (\mathbf{r}) = V_{AA'} (\mathbf{r} + (\mathbf{L}_+ + \mathbf{L}_-)/3)$, 
$V_{BA'} (\mathbf{r}) = V_{AA'} (\mathbf{r} - (\mathbf{L}_+ + \mathbf{L}_-)/3)$.
A common procedure is to assume that the interlayer hopping is dominated by
processes with momentum-transfer $\mathbf{Q}_0 = 0$ or equal to the reciprocal
vectors $\mathbf{Q}_{1,2}$, so that 
$V_{AA'} (\mathbf{r}) \approx 
 (w/v_F) \sum_j \exp (i \mathbf{Q}_j \cdot \mathbf{r} )$ \cite{port,bist}.
For the lowest subband obtained using this approximation, the energy contour 
lines corresponding to the bilayers with $n = 10$ and 22 in the sequence 
(\ref{seq}) are represented in Figs. \ref{vhtwist}(a) and \ref{vhtwist}(b). 
In the plots, it is manifest the 
development of saddle points in the dispersion between each pair of neighboring 
Dirac points, although with a little lateral displacement that breaks the 
symmetry down to $C_{3v}$.

{\em Pairing instabilities.---} When analyzing the possible pairing 
instabilities in this system, one has to pay attention to the fact that the 
Fermi line of the twisted bilayer is made of two disconnected sections, which 
are related by the inversion ${\bf k} \rightarrow -{\bf k}$ as represented in 
Fig. \ref{vhtwist}(c). This means that we can 
distinguish between two different processes in the scattering of the Cooper 
pairs, depending on whether there is exchange or not of the two electrons from 
one section of the Fermi line to the other. The strength of the Coulomb 
interaction is different in the two cases, as the exchange of the electrons in 
the Cooper pair implies a large momentum transfer of the order of $|{\bf K}|$, 
while that must be of order $\sim 1/L$ when each electron remains in the same 
section of the Fermi line. In the first instance, the Coulomb repulsion is 
given by the unscreened potential $v_0 ({\bf K}) = 2\pi e^2 /|{\bf K}|$. In 
the other case, the interaction is already screened for momenta $\sim 1/L$. 
This effect can be estimated from the charge density required to place the 
Fermi level close to the Van Hove singularity, located at an energy 
$\varepsilon_{\rm VHS} \sim v_F/L$ from the Dirac point. Applying the charge 
polarization $\chi_D ({\bf k})$ in the Dirac theory for a number of $N_D = 8$ 
doped Dirac cones (including spin) in the twisted bilayer 
\begin{equation}
\chi_D \left( \tfrac{1}{L} \right) 
 \approx \frac{N_D \: \varepsilon_{\rm VHS}}{2\pi v_F^2}  \sim  \frac{N_D}{2\pi v_F L}
\end{equation} 
we can estimate the strength of the screened Coulomb potential as
\begin{eqnarray}
v\left( \tfrac{1}{L} \right)  & = &  \frac{v_0 \left( \tfrac{1}{L} \right)}
   {1 + v_0 \left( \tfrac{1}{L} \right)  \chi_D \left( \tfrac{1}{L} \right)}
                                                       \nonumber        \\
  &  \sim  &   \frac{2\pi v_F L}{ N_D}
\end{eqnarray}

The important point is that, already for a period $L_n$ with $n \approx 10$, 
the strength of the effective interaction $v(1/L_n)$ is about one order of 
magnitude above that of the unscreened potential $v_0 ({\bf K})$. Then, it is 
safe to neglect the influence of the scattering of 
Cooper pairs with large momentum transfer in the evaluation of the initial 
condition (\ref{init}). Moreover, with the system placed in close proximity
to the Van Hove singularity, the electron-hole susceptibility only experiences
a significant enhancement at small momentum. Therefore, in order to study 
the effect of the singularities at the right-hand-side of Eq. (\ref{init}), it 
is justified to discard the second term in favor of the third, which is the
only source of a possible pole.

We observe that the previous argument leading to a negative coupling $V_{3,3}$
from the relation (\ref{v33}) does not apply in the case of the twisted graphene
bilayers, as it was based on the dominance of the term depending on 
$\chi_{\rm ph}(\mathbf{k}+\mathbf{k}^\prime)$ in Eq. (\ref{init}) for monolayer
graphene. In the present case, one can still decompose the BCS vertex
as a series of the basis functions for irreducible representations of the 
$C_{3v}$ point symmetry group. Two of these are one-dimensional, corresponding 
to the sets $\{ \cos ((3n)\theta) \}$ and $\{ \sin ((3n)\theta) \}$ for integer 
$n$, and the remaining is two-dimensional, represented by the set 
$\{ \cos (m \theta) , \sin (m \theta) \}$ for integer $m$ not being a multiple 
of 3. It has to be remarked, however, that the electron-hole susceptibility 
lacks in the twisted bilayers the very large enhancement at vanishing momentum 
that was the consequence of the extended character of the Van Hove singularity 
in monolayer graphene. This does not prevent that some negative coupling
may arise in any of the channels for the above representations, but
one can anticipate that the effect will not correspond now to the dominance 
of a particular harmonic from the set of basis functions.

\begin{figure}[t]
\begin{center}
\mbox{\epsfxsize 6.0cm \epsfbox{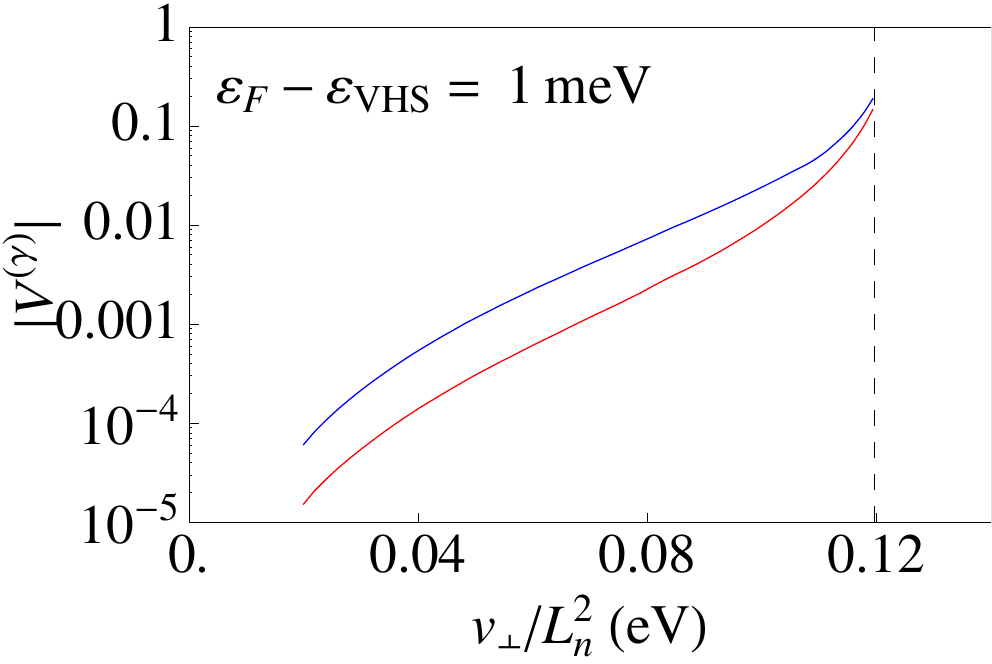}}
\end{center}
\caption{Plot of the absolute value of the negative couplings $V^{(\gamma )}$ 
for the irreducible representations $\{ \sin ((3n)\theta) \}$ (upper curve)
and $\{ \cos ((3n)\theta) \}$ (lower curve) of the $C_{3v}$ group, as  
functions of the ratio of the potential $v_\perp$ to the square of the 
lattice constant $L_n$ (for a twisted bilayer with $n = 10$). The dashed line
represents the point of the singularity in the spin response function for the
same value of $L_n$.}
\label{evotw}
\end{figure}

To estimate the scale of a possible pairing instability, we have computed the 
polarization $\chi_{\rm ph}(\mathbf{q})$ in a grid with $600 \times 600$ 
points in each hexagon of the the Brillouin zone, with the Fermi line close to 
the saddle points of the dispersions represented in Fig. \ref{vhtwist}. A 
characteristic plot of
the couplings in the channels with effective attractive interaction is shown 
in Fig. \ref{evotw}. We observe that the magnitude of the couplings is 
significantly smaller than that found for the Van Hove singularity in the 
conduction band of graphene. Taking into account the correspondence 
(\ref{ctemp}) with the critical scale of a possible pairing instability, it is 
clear that the present effective attraction is too small to give rise to any 
observable effect in the twisted bilayers. More importantly, the couplings do 
not grow large when approaching the singularity in the second fraction of 
Eq. (\ref{init}), which means that the corresponding divergence 
becomes integrable when computing the coefficients in the expansion 
(\ref{dec}). This implies that the tendency towards pairing cannot compete in 
this case with the instability in the spin response function, no matter how 
close the Van Hove singularity is approached.

\begin{figure}[t]
\begin{center}
\mbox{\epsfxsize 5.5cm \epsfbox{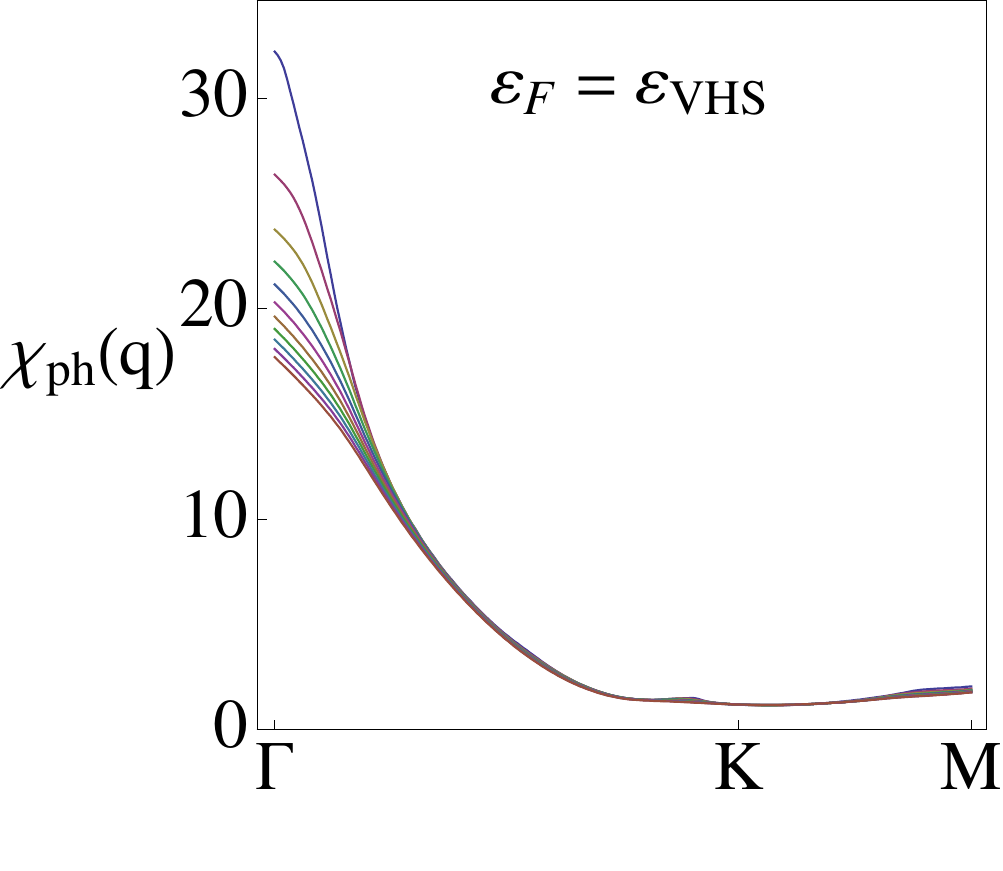}}
\hspace{1cm}
\mbox{\epsfxsize 5.0cm \epsfbox{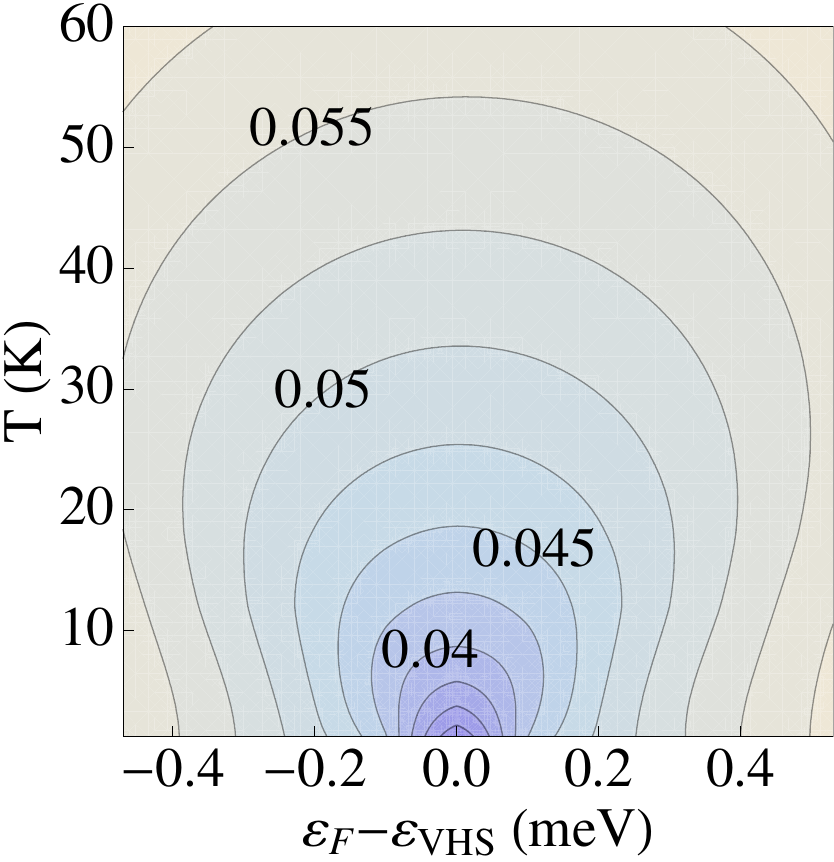}}
\end{center}
\hspace{0.9cm}  (a)  \hspace{5.5cm}  (b)
\caption{(a) Plot of the electron-hole susceptibility (in units of the inverse 
of eV times $L_n^2$) for temperatures corresponding (from top to bottom) to 
$k_B T$ = 0.1, 0.5, 1, 1.5, 2, 2.5, 3, 3.5, 4, 4.5, 5 meV, and (b) contour 
lines of the critical coupling for the interaction strength $v_{\perp}/L_n^2$ 
(in eV) as a function of temperature and doping with respect to the Van Hove 
singularity, for a twisted bilayer with $n = 10$ in the sequence (\ref{seq}).}
\label{ferro}
\end{figure}

\begin{figure}[h]
\begin{center}
\mbox{\epsfxsize 5.5cm \epsfbox{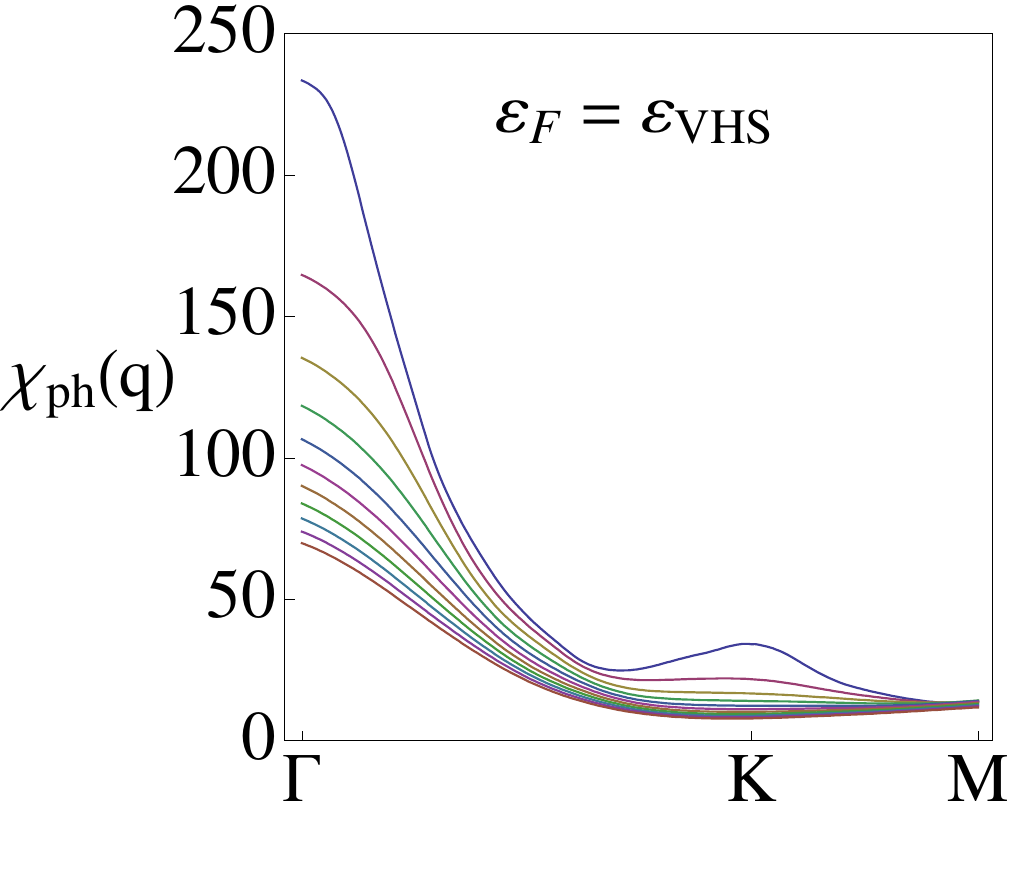}}
\hspace{1cm}
\mbox{\epsfxsize 5.0cm \epsfbox{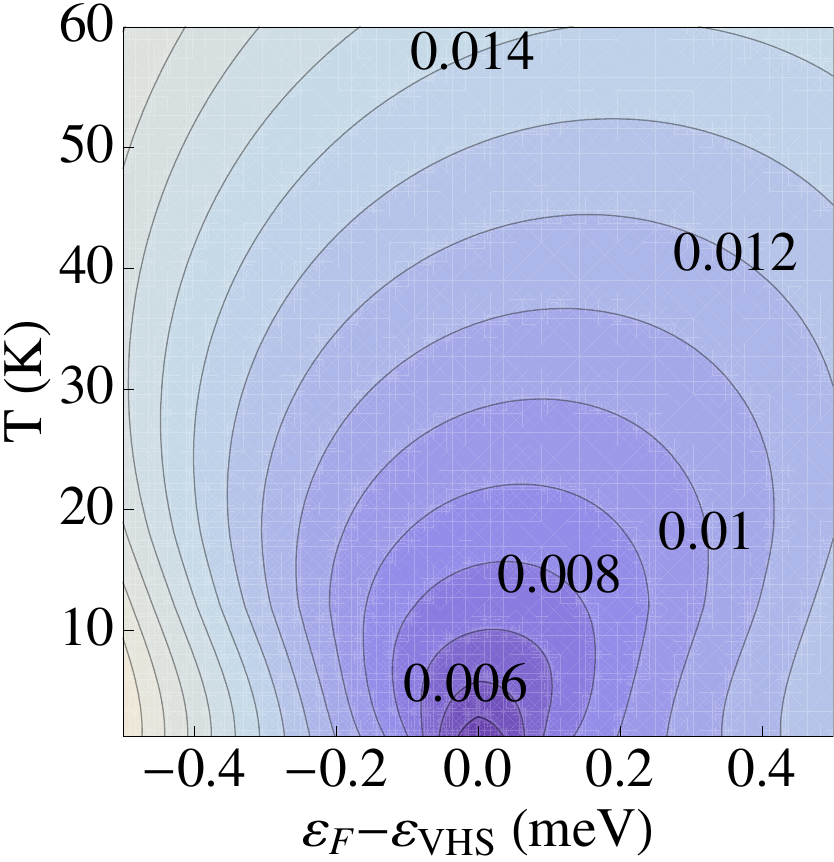}}
\end{center}
\hspace{1.1cm}  (a)  \hspace{5.5cm}  (b)
\caption{(a) Plot of the electron-hole susceptibility (in units of the inverse 
of eV times $L_n^2$) for temperatures corresponding (from top to bottom) to 
$k_B T$ = 0.1, 0.5, 1, 1.5, 2, 2.5, 3, 3.5, 4, 4.5, 5 meV, and (b) contour 
lines of the critical coupling for the interaction strength $v_{\perp}/L_n^2$ 
(in eV) as a function of temperature and doping with respect to the Van Hove 
singularity, for a twisted bilayer with $n = 22$ in the sequence (\ref{seq}).}
\label{ferro22}
\end{figure}

{\em Ferromagnetic instability.---} In order to assess the strength of the 
magnetic instability in the system, we have evaluated 
$\chi_{\rm ph}(\mathbf{q})$ along high-symmetry directions down to vanishing 
momentum, using this time a grid with $900 \times 900$ points covering one of 
the hexagons of the Brillouin zone. Computing at decreasing temperatures and 
doping levels progressively close to the Van Hove singularity, we observe the 
development of a pronounced peak in the susceptibility at zero momentum, as 
shown in Figs. \ref{ferro}(a) and \ref{ferro22}(a). This is consistent with the 
logarithmic divergence anticipated in the continuum approach of Sec. II. As 
seen in that section, such a behavior implies that, ideally, the singularity
in the response function $R_s ({\bf 0},\omega )$ can be reached for any 
strength of the local Coulomb repulsion, simply by lowering the temperature and 
approaching sufficiently close the Van Hove singularity.

We remark that, in the model of the bilayer superlattice, the  
spin-dependent interaction $v_{\perp}$ arises at a microscopic level 
from the on-site Coulomb repulsion $U$ at each carbon atom. The eigenstates 
in each subband of the twisted bilayer can be written as linear 
combinations of the atomic orbitals in the superlattice unit cell, normalized 
according to the number $M$ of atoms it contains. This means that, after 
projecting into the lowest-energy subband,  $v_{\perp}/L_n^2$ must be of the 
order of $U/M$, having a magnitude that decreases in the sequence of twisted 
bilayers as the inverse of $L_n^2$. That has to be confronted with the critical 
values that are predicted from the position of the pole in the response 
function $R_s ({\bf 0},\omega )$, which we have represented for a twisted 
bilayer with $n = 10$ in Fig. \ref{ferro}(b). Assuming a magnitude of 
the on-site Coulomb repulsion $U \sim 10$ eV \cite{kat}, we obtain 
$U/M \sim  0.03$ eV for that particular superlattice. We see that such an 
interaction strength cannot match the lowest critical couplings represented 
in the figure, staying in close proximity to the Van Hove singularity. Thus, 
for the corresponding twisted bilayer, a signature of the magnetic instability 
should be only expected at a temperature  $\lesssim 1$ K.

We have anyhow to bear in mind that the band structure of the twisted bilayers
undergoes important changes as the period $L_n$ increases, with a progressive
reduction in the width of the lowest-energy subband. It has been actually found 
that such a bandwidth has a recurrent behavior, narrowing down to approximately 
zero energy at a sequence of ^^ ^^ magic" twist angles\cite{bist,prl3}. The 
first instance at which this happens corresponds to $n = 31$. Then, in the way 
towards this magic angle, the lowest subband of the bilayer becomes 
increasingly flat, which has a significant impact on the values of the
electron-hole  
susceptibility. This can be appreciated in the plot of Fig. \ref{ferro22}(a), 
which shows the result of computing $\chi_{\rm ph}(\mathbf{q})$ for 
a twisted bilayer with $n = 22$. From the position of the pole in the spin 
response function $R_s ({\bf 0},\omega )$, one can calculate again the critical 
values of the interaction strength needed to reach the magnetic instability. 
These are represented in Fig. \ref{ferro22}(b) as a function of the temperature 
and the deviation of the filling level with respect to the Van Hove 
singularity. With a rough estimate of $U/M \sim  0.007$ eV for the 
corresponding twisted bilayer, we find that the onset of the instability may 
now take place at temperatures approaching the order of magnitude of $10$ K.

While the ferromagnetic instability is increasingly amplified for twisted 
bilayers becoming closer to the first magic angle at $n = 31$, it is clear that 
our computational approach must become at some point unreliable to estimate 
quantitatively the critical interaction strength. As mentioned before, 
the framework described in Sec. IIA leads to results similar to those obtained 
in a renormalization group approach to the magnetic instability. From this 
point of view, the present computational scheme is justified whenever the 
interaction strength is much smaller than the bandwidth of the 
electron system. This certainly happens in the two cases considered with
$n = 10$ and 22. The scaling approach must break down however in twisted 
bilayers for some larger lattice constant as the bandwidth goes to zero, 
leaving then ferromagnetism as a likely instability but much harder to 
analyze quantitatively.

\section{Conclusions}

In this paper we have studied the many-body instabilities of electrons 
interacting near Van Hove singularities arising in monolayer and twisted 
bilayer graphene. In the first instance, we have taken advantage of the 
experimental data available for the dispersion around the saddle points 
in the conduction band of graphene\cite{prl}. While reaching the level of the 
singularity requires there a large amount of doping, the results obtained 
from ARPES have unveiled the extended character of the saddle point 
dispersion, showing the potential for a large instability of the electron 
system. On the other hand, twisted graphene bilayers have Van Hove 
singularities that arise from the hybridization of the Dirac cones of the 
two layers, being therefore located at relatively low energies from the 
charge neutrality point. They can be more easily reached upon doping, 
but have the disadvantage of involving in general a weaker singularity in 
the density of states, with lower strength of the consequent electronic 
instabilities.

We have seen that a pairing instability must be dominant over the tendency 
to magnetic order as the Fermi level is tuned to the Van Hove singularity in 
the conduction band of graphene. As a result of the extended character of 
the saddle points in the dispersion, we have found that the pairing of the 
electrons must take place preferentially in a channel of $f$-wave symmetry, 
with an order parameter vanishing at the position of the saddle points 
along the Fermi line. In the case of the twisted bilayers, the dispersion 
has instead its symmetry reduced down to the $C_{3v}$ group and, most 
importantly, it leads to susceptibilities that diverge at the saddle points 
but are integrable along the Fermi line. This implies that the attractive 
couplings for the pairing instability do not grow large when approaching the 
critical interaction strength marking the onset of magnetic order. Thus,
the magnetic instability becomes prevalent in the twisted graphene bilayers, 
with a dominant tendency towards ferromagnetism as the uniform magnetic 
susceptibility inherits the divergence in the density of states at the Van 
Hove singularity.

The divergence of the electron-hole susceptibility is the driving force for 
pairing and magnetic instabilities at the Van Hove singularity. In principle,
such a susceptibility can be made arbitrarily large in the limit of vanishing
temperature. In practice, however, there is a restriction to that growth 
coming from the effects of disorder or defects in the carbon lattice. These 
have the ability to smear the singularity in the density of states, reducing
its nominal strength. The most dangerous effect comes in that respect 
from the scattering off impurities, and the analysis made in the two-dimensional 
square lattice has shown that the logarithmic singularity in the density of 
states is suppressed by corrections that depend on the function 
$(1/4\pi t\tau)\log (4\tau |\varepsilon |)$, $t$ being the nearest-neighbor 
hopping amplitude and $\tau $ the relaxation time\cite{turk}. In our case, for 
sufficiently 
clean samples, we can assume for instance a value of the mean free path in the 
carbon lattice of the order of $\sim 1 \mu$m, which would imply a relaxation 
time $\tau \sim 10^4 t^{-1}$. Thus, we see that the divergence in the density 
of states may not suffer a significant attenuation when probed with a 
resolution of $\sim 0.1$ meV. This level of approximation to the singularity 
may be small enough to observe the onset of the electronic instabilities since, 
as shown above, it corresponds to doping levels at which experimental 
signatures can be seen for reasonable values of the interaction strength in 
monolayer as well as in twisted bilayer graphene.

The results presented in this paper may serve as a guide for the effects
that can be measured in real graphene samples near a Van Hove singularity.
Renormalization group methods have shown to provide in general a predictable 
approach to the low-energy physics of the singular density of states, where 
effects like the quasiparticle attenuation may be compensated in part by the 
renormalization of the saddle point dispersion\cite{np}. The carbon-based 
systems studied in the present paper may provide an appropriate playground to 
test many of the physical effects predicted, including the regime of strong 
correlations supposed to arise in the vicinity of the Van Hove singularity.

\section*{Aknowledgments}

I thank P. San-Jos\'e for his support at the early stages of the work on 
twisted graphene bilayers. I also thank F. Guinea for very useful conversations.
The financial support from MICINN (Spain) through grant FIS2011-23713 is 
gratefully acknowledged.

\end{document}